\documentclass[iop,twocolappendix]{emulateapj}

\usepackage{graphicx}
\usepackage{natbib}
\usepackage{url}
\usepackage{multirow}
\usepackage{amsmath}
\usepackage{subfig}
\usepackage{textcomp}
\usepackage{caption}
\usepackage{booktabs}

\renewcommand{\captionsize}{\footnotesize}

\newcommand{\Hersc}{{\it Herschel}}

\newcommand{\IRAS}{{\it IRAS\/}}
\newcommand{\ISO}{{\it ISO\/}}
\newcommand{\Spitzer}{{\it Spitzer\/}}
\newcommand{\akari}{{\it AKARI\/}}

\shorttitle{HELGA II: Dust and Gas in Andromeda}
\shortauthors{Smith, Eales, Gomez et al.}

\begin{document}

\title{The Herschel Exploitation of Local Galaxy Andromeda (HELGA) II: Dust and Gas in Andromeda
       \footnotemark[*]}

\submitted{Accepted to ApJ July 2012}

\author{M. W. L. Smith\altaffilmark{1},
        S. A. Eales\altaffilmark{1},
        H. L. Gomez\altaffilmark{1},
	J. Roman Duval\altaffilmark{2},
	J. Fritz\altaffilmark{3},
	R. Braun\altaffilmark{4},
        M. Baes\altaffilmark{3},
        G. J. Bendo\altaffilmark{5}, 
	J. A. D. L. Blommaert\altaffilmark{6},
	M. Boquien\altaffilmark{7},
	A. Boselli\altaffilmark{7},
        D. L. Clements\altaffilmark{8},
	A. R. Cooray\altaffilmark{9},
	L. Cortese\altaffilmark{10},
        I. de Looze\altaffilmark{3},
        G. P. Ford\altaffilmark{1},
	W. K. Gear\altaffilmark{1},
	G. Gentile\altaffilmark{3},
        K. D. Gordon\altaffilmark{2,3},
        J. Kirk\altaffilmark{1},
	V. Lebouteiller\altaffilmark{11},
        S. Madden\altaffilmark{11},
	E. Mentuch\altaffilmark{12},
        B. O'Halloran\altaffilmark{8},
        M. J. Page\altaffilmark{13},
        B. Schulz\altaffilmark{14},
	L. Spinoglio\altaffilmark{15},
        J. Verstappen\altaffilmark{3},
        C. D. Wilson\altaffilmark{16}}

\altaffiltext{1}{School of Physics \& Astronomy, Cardiff University,
  The Parade, Cardiff, CF24 3AA, UK}
  \email{Matthew.Smith@astro.cf.ac.uk}
\altaffiltext{2}{Space Telescope Science Institute, 3700 San Martin Drive, Baltimore, MD 21218}
\altaffiltext{3}{Sterrenkundig Observatorium, Universiteit Gent, Krijgslaan 281 S9, B-9000 Gent, Belgium}
\altaffiltext{4}{CSIRO Astronomy and Space Science, P.O. Box 76, Epping, NSW 1710, Australia}
\altaffiltext{5}{UK ALMA Regional Centre Node, Jodrell Bank Centre for Astrophysics, School of Physics and Astronomy, University of Manchester, Oxford Road, Manchester, M13 9PL, United Kingdom}
\altaffiltext{6}{Instituut voor Sterrenkunde, K.U.Leuven, Celestijnenlaan 200D, B-3001 Leuven, Belgium}
\altaffiltext{7}{Laboratoire d'Astrophysique de Marseille - LAM, Université d'Aix-Marseille \& CNRS, UMR7326, 38 rue F. Joliot-Curie, 13388, Marseille Cedex 13, France}
\altaffiltext{8}{Astrophysics Group, Imperial College, Blackett Laboratory, Prince Consort Road, London SW7 2AZ, United Kingdom}
\altaffiltext{9}{Center for Cosmology and the Department of Physics \& Astronomy, University of California, Irvine, CA 92697, USA}
\altaffiltext{10}{European Southern Observatory, Karl Schwarschild Str. 2, 85748, Garching bei Muenchen, Germany}
\altaffiltext{11}{CEA, Laboratoire AIM, Irfu/SAp, Orme des Merisiers, F-91191, Gif-sur-Yvette, France}
\altaffiltext{12}{Astronomy Department, University of Texas at Austin, Austin, TX 78712, USA}
\altaffiltext{13}{Mullard Space Science Laboratory, University College London, Holmbury St. Mary, Dorking, Surrey RH5 6NT, UK}
\altaffiltext{14}{Infrared Processing and Analysis Center, MS 100-22, California Institute of Technology, JPL, Pasadena, CA 91125, USA}
\altaffiltext{15}{Istituto di Fisica dello Spazio Interplanetario, INAF, via del Fosso del Cavaliere 100, 00133 Roma, Italy}
\altaffiltext{16}{Department of Physics \& Astronomy, McMaster University, Hamilton, ON, L8S 4M1, Canada}

\begin{abstract}

We present an analysis of the dust and gas in Andromeda, using \Hersc\ images sampling the entire far-infrared peak.
We fit a modified-blackbody model to $\sim$4000 quasi-independent pixels with spatial resolution of $\sim$140\,pc and
find that a variable dust-emissivity index ($\beta$) is required to fit the data. We find no significant long-wavelength excess above this model 
suggesting there is no cold dust component. We show that the gas-to-dust ratio  
varies radially, increasing from $\sim$20 in the center to $\sim$70 in the star-forming ring at 10\,kpc, consistent with the 
metallicity gradient. In the 10\,kpc ring the average $\beta$ is $\sim$1.9, in good agreement with values determined for the Milky Way (MW).
However, in contrast to the MW, we find significant radial variations in $\beta$, which increases from 1.9 at 10\,kpc to $\sim$2.5 at a radius of 3.1\,kpc 
and then decreases to 1.7 in the center. The dust temperature is fairly constant in the 10\,kpc ring (ranging from 17--20\,K), but increases
strongly in the bulge to $\sim$30\,K. Within 3.1\,kpc we find
the dust temperature is highly correlated with the 3.6\micron\ flux, suggesting the general stellar population in the bulge is the dominant source of dust heating there. 
At larger radii, there is a weak correlation between the star formation rate and dust temperature. We find no evidence for `dark gas' in M31 in contrast to recent results for the MW. 
Finally, we obtained an estimate of the CO X-factor by minimising the dispersion in the gas-to-dust ratio, obtaining a value of 
$(1.9 \pm 0.4)\times 10^{20}$\,$\rm cm^{-2} [K\,kms^{-1}]^{-1}$.

\end{abstract}


\section{Introduction}
\label{sec:intro}

Astronomy at long infrared wavelengths (20--1000\micron) is a relatively young field due to the need for space missions to avoid the absorption of the atmosphere in this waveband.
This waveband, however, is vital for astronomical studies as this is where dust in the interstellar medium (ISM) radiates. 
This is important for studies of galaxy evolution
as star formation regions tend to be dusty, and therefore the use of UV and optical measurements to trace the star formation rate
can lead to it being underestimated
\citep[see][]{Kennicutt1998, Blain1999, Calzetti2001, Papovich2002, Calzetti2010}.
Calibrating the relationship between infrared emission and star formation 
rate has been difficult due to uncertainties from the contribution of the general stellar population to heating the dust, the fact that not all optical/UV emission is absorbed, and uncertainties in 
the properties of the dust.
The dust emission is affected by the composition of the dust, and the proportion of non-equilibrium to equilibrium heating. Studies with previous 
space missions \IRAS, \ISO, \Spitzer\ and \akari\ tried to address these questions \citep[e.g.,][]{Walterbos1996,Boselli2004,Calzetti2010,Buat2011}. However, as they were limited to wavelengths
less than 160\micron, they were not sensitive to the cold dust ($\leq$\,15\,K) and missed up to 50\% of the dust mass in galaxies.

The continuum emission from the dust has been  
{\footnotesize \\ \\ $^*$\textit{H\lowercase{erschel}} \lowercase{is an} ESA \lowercase{space observatory 
		 with science instruments provided by} E\lowercase{uropean-led} P\lowercase{rincipal}
		I\lowercase{nvestigator consortia and with important participation from} NASA.\\}
proposed as a potential method of measuring the total 
mass of the ISM \citep{Hildebrand1983, Guelin1993, Eales2010, Eales2012}; 
traditionally the amount of gas has been measured using H{\sc i} 
and CO surveys, but due to sensitivity
and resolution issues this method is limited to low redshift and small numbers of galaxies. 
\citet{Smith2012} found that for early-type galaxies (E/S0) the ISM was detected for 50\% of objects through its dust emission but
only 22\% through its CO emission.
In addition, the conversion of the CO tracer to molecular gas mass is highly uncertain and is believed to vary with metallicity 
\citep[see][]{Wilson1995, Boselli2002, Strong2004, Israel2005, Narayanan2012}. This is a topical area as recent studies using
the Planck all-sky survey and \IRAS\ maps have made an estimate of the amount of `dark gas' \citep{Planck2011} in the Milky Way. 
The `dark gas' is thought to be molecular gas which is traced by dust, but not detected with the standard CO method. Previous works have also suggested
the presence of `dark gas' by using $\gamma$-ray emission from cosmic-ray interactions with clouds of gas \citep{Grenier2005,Abdo2010} and by the kinematics of recycled dwarf galaxies \citep{Bournaud2007}.

\Hersc\ is one of the European Space Agency's flagship observatories and observes in the far-infrared (FIR) in the range 55--671\micron\ \citep[see][]{Pilbratt2010}. Due to the large space mirror and cryogenic 
instruments, it has a high sensitivity and unprecedented angular resolution at these wavelengths. \Hersc\ has the ability to target both large numbers of galaxies and map large areas of sky. 
It has two photometric instruments: PACS \citep{Poglitsch2010} which can observe in 3 broad bands around 70, 100 and 160\micron\ (70 and 100\micron\ cannot be used simultaneously) and
SPIRE \citep{Griffin2010} which observes simultaneously in filter bands centered at 250, 350 and 500\micron. SPIRE provides flux measurements 
on the longer wavelength side of the FIR dust peak ($\sim$160\micron), allowing us to obtain a full census of dust in nearby galaxies \citep[e.g.,][]{Dunne2011}.

Andromeda (M31) and the Milky Way (MW) are the only two large spirals in the Local Group. Studies of Andromeda therefore provide
the best comparison to observations of our own Galaxy with the advantage that we get a `global' picture, whereas investigations of the Milky Way
are limited by our location within the Galaxy. The total size of M31 and the scalelength of its disk are both approximately twice those of the MW \citep[see][and references therein]{Yin2009}.
However, the star-formation rate of the MW is $\sim$3--6\,$M_{\odot}\rm\,yr^{-1}$ \citep{Boissier1999} compared to only $\sim$0.3--1.0 in M31
\citep[][G. P. Ford et al., in preparation]{Barmby2006,Williams2003}, despite similar amounts of gas present \citep{Yin2009}. For this reason, M31 is often labelled
as `quiescent'. The dust emission from M31 is dominated by a dusty star-forming ring at a radius of 10\,kpc, and was first observed in the infrared by \IRAS\ \citep{Habing1984}.

Many projects to map the ISM in M31, have been undertaken using observations in the mid-infrared (MIR) with \Spitzer\ \citep{Barmby2006}, 
in the FIR with \Spitzer\ \citep{Gordon2006}, in the H{\sc i} atomic line \citep{Thilker2004, Braun2009, Chemin2009} 
and in the CO($J$=1-0) line \citep{Nieten2006}. Studies of the gas kinematics and dust emission show M31 has a 
complicated structure \citep[e.g.,][]{Chemin2009, Gordon2006}; \citet{Block2006} attribute many of the features observed to density waves from a 
possible head-on collision with M32. \citet{Tabatabaei2010} investigated the relation between gas, dust and star formation using FIR \Spitzer\ data.
\citet{Leroy2011} used \Spitzer\ and gas observations to investigate the conversion between CO($J$=1-0) line flux and molecular hydrogen column-density 
in a sample of nearby galaxies including M31. In their analysis they
found M31 has the lowest dust temperatures in their sample and therefore would benefit most by including \Hersc\ data.

The \Hersc\ Exploitation of Local Galaxy Andromeda (HELGA) is a survey covering a $\sim 5.5^{\circ}\times 2.5^{\circ}$ area centered on M31. We use PACS-SPIRE parallel mode,
observing at 100, 160, 250, 350 and 500\micron\ simultaneously. Further details of the observations can be found in Section \ref{sec:FIRobs} and in \citet{Fritz2011}.
HELGA observations have been used in other works to investigate structures in dust and H{\sc i} at large radii \citep{Fritz2011}, 
the relationship between gas and star formation (G. P. Ford et al., in preparation),
and the structure of M31 and the cloud-mass function (J. Kirk et al., in preparation).

In this paper we use the \Hersc\ data combined with the wealth of ancillary data to investigate the distribution, 
emission properties and the processes heating the dust in M31 on spatial scales of $\sim$140\,kpc.
There have still been relatively few attempts to map the dust within a galaxy; 
recent studies with \Hersc\ include \citet{Smith2010}, \citet{Boquien2011}, \citet{Foyle2012}, \citet{Bendo2012}, and
\citet{Mentuch2012}, but they are often limited to small numbers of independent pixels. 
The advantage of M31 over these other studies is the close proximity which allows us to investigate
the dust at higher spatial scale and with many more independent pixels.
We also apply the \textit{Planck} method for finding `dark gas' \citep{Planck2011} to M31 and use this method to determine a value of the X-factor ---
the relationship between the molecular hydrogen column density and the observed CO tracer.
In Section \ref{sec:data} we present
the data used for this analysis and Section \ref{sec:FIRsed} describes our method for fitting the spectral energy distribution of dust.
Section \ref{sec:res} presents our maps of the dust properties, including the distribution of dust surface density, temperature and spectral index.
This section also describes a comparison of the distributions of gas and dust
and of the search for an excess emission at long wavelengths. In Section \ref{sec:disc} we discuss the 
dust properties including the composition of the dust and the processes heating the dust. 
In this section we also search for `dark gas' and determine the value of the X-factor. 
The conclusions are presented in Section \ref{sec:conc}.

\section{The Data}
\label{sec:data}

\subsection{Far-Infrared Observations}
\label{sec:FIRobs}

\Hersc\ observations of M31 were taken using both PACS and SPIRE in parallel mode covering an area of $\sim 5.5^{\circ} \times 2.5^{\circ}$ centered
on M31. To observe an area this large, the observations were split into two halves with a cross-scan on each half, 
which produced data at 100, 160, 250, 350 and 500\micron\ simultaneously (observation IDs: 1342211294, 1342211309, 1342211319, 1342213207). 
Full details of the observing strategy and data reduction can be found in \citet{Fritz2011}. 

The PACS data reduction was performed in two stages. The initial processing up to Level-1 (i.e., to the level where
the pointed photometer timelines have been calibrated) is performed in HIPE v8.0 
\citep{Ott2010} using the standard pipeline. To remove low-frequency noise (or drifts) in the arrays, residual glitches and create the final maps we use SCANAMORPHOS \citep[v15,][]{Roussel2011}. 
The final maps have a pixel scale of 2\arcsec\ and 3\arcsec\ and a spatial resolution of 12.5\arcsec\ and 13.3\arcsec\ full-width half maximum (FWHM) at 100 and 160\micron\ respectively. 

\begin{figure*}
  \centering
  \includegraphics[trim=64.0mm 32.0mm 1.0mm 4.0mm,clip=true,width=0.88\textwidth]{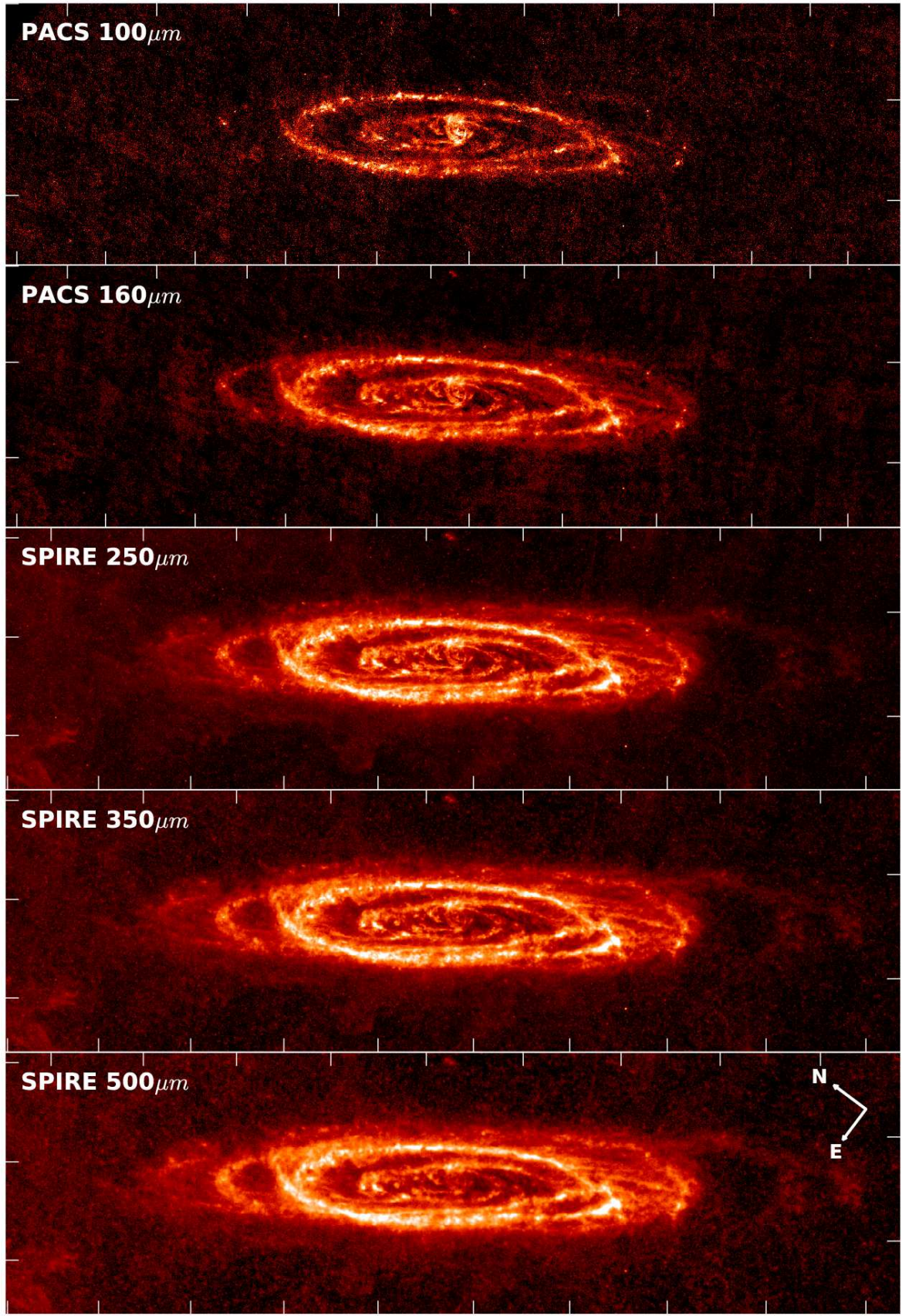}
  \caption{\Hersc\ images used in the analysis of this paper. The images have dimensions of approximately $\rm 4.5^{\circ}\times 1.3^{\circ}$,
           with a tick spacing of $30^{\prime}$ and centered on $10^{h} 43^{m} 02^{s}$, $+41^{\circ} 17^{\prime} 42^{\prime\prime}$. These images
           are all at their original resolution.}
  \label{fig:her-maps}
\end{figure*}

The SPIRE data were processed up to Level-1 with a custom pipeline script adapted from the official pipeline. 
We apply the latest flux correction factors from the SPIRE Instrument Control Centre to update the maps to the latest calibration product \citep{SPIREOM}. 
For the baseline subtraction we use a custom method called {\sc BriGAdE} (M. W. L. Smith et al., in preparation) which uses an alternative technique for correcting temperature drifts. The final maps
were created using the naive mapper with pixel sizes of 6\arcsec, 8\arcsec and 12\arcsec\ with spatial resolution of 18.2\arcsec, 24.5\arcsec, 36.0\arcsec\ FWHM for the 250, 350 and 500\micron\ maps, respectively.
All \Hersc\ images are shown in Figure \ref{fig:her-maps}.

In addition to the \Hersc\ data, we make use of the 70\micron\ \Spitzer\ MIPS \citep{Rieke2004} map published in \citet{Gordon2006}. 
This observation covers a region of M31 approximately $3^{\circ} \times 1^{\circ}$
in size and has an exposure time of $\sim 40$ s pixel$^{-1}$. The data was processed using the MIPS Data Analysis Tool, version 2.90 \citep{Gordon2005} and full details of the reduction
can be found in \citet{Gordon2006}.

\subsection{Gas Measurements}

To investigate the atomic hydrogen in Andromeda we use the H{\sc i} moment-zero map presented in 
\citet{Braun2009}. The observations were made with the Westerbork Synthesis Radio Telescope (WRST)
covering a region of $\sim 6^{\circ} \times 3.5^{\circ}$ with a resolution of 18\arcsec $\times$ 15\arcsec. 
In this work we present our results using a map which has not been corrected for opacity effects since this correction is uncertain. 
For the results which make use of the H{\sc i} map we also test how our results are affected by using a H{\sc i} map corrected for self-opacity using the prescription
outlined in \citet{Braun2009}. The uncorrected H{\sc i} map has a sensitivity of $\rm 4.2\times 10^{18} cm^{-2}$. 

For the molecular hydrogen gas content we use CO($J$=1-0) observations presented in \citet{Nieten2006} made with the IRAM 30m telescope. 
This covers an area of $2^{\circ} \times 0.5^{\circ}$ with a sensitivity of $\sim$0.35\,K\,km\,s$^{-1}$.

\begin{figure*}
  \centering
  \includegraphics[trim=48.0mm 25.0mm 1.0mm 4.0mm,clip=true,width=0.88\textwidth]{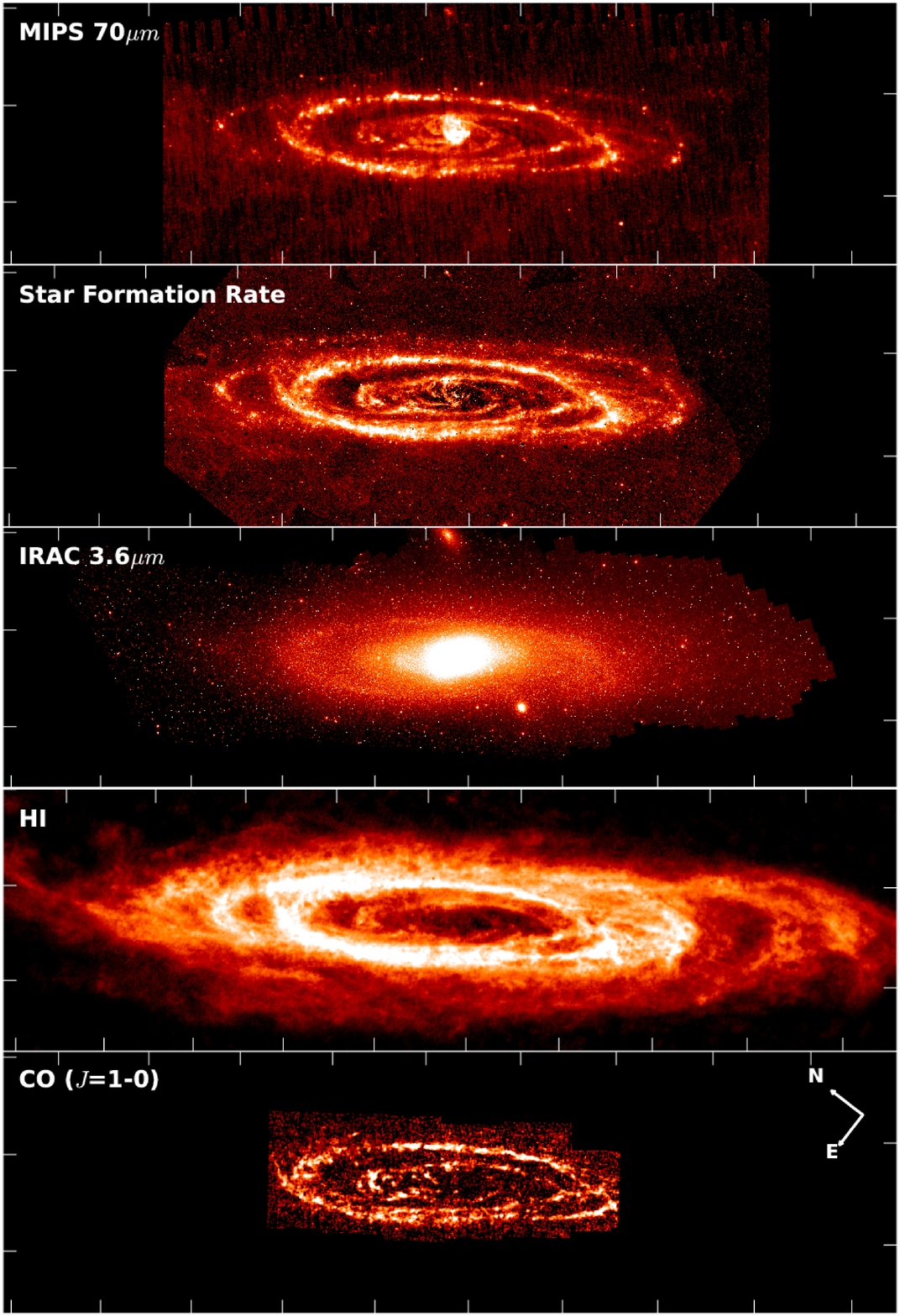}
  \caption{Ancillary images for M31. The scale is the same as for the Herschel images presented in Figure \ref{fig:her-maps}. From top:
           \Spitzer\ MIPS 70\micron, the star formation rate (presented in G. P. Ford et al. in preparation), \Spitzer\ IRAC 3.6\micron\ (presented in \citealt{Barmby2006}), 
           H{\sc i} and CO images as in Figure \ref{fig:her-maps}. The CO map (used as a tracer of $\rm H_{2}$) only covers an area of 
           $\rm 2^{\circ} \times 0.5^{\circ}$. These images are all at their original resolution.}
  \label{fig:oth-maps}
\end{figure*}

All maps other than the \Hersc\ images used for this analysis are shown in Figure \ref{fig:oth-maps}.

\section{The FIR-Submm Spectral Energy Distribution}
\label{sec:FIRsed}

To investigate how the dust properties vary across M31 we undertake a pixel-by-pixel 
dust analysis, using the \Spitzer\ 70\micron, \Hersc\ PACS and SPIRE images. We first convolve the data to the same resolution as the 500\micron\
map (our largest FWHM) by using a kernel which matches the point-spread function (PSF) in a particular band to the 500\micron\ band. 
This procedure is described in detail in \citet{Bendo2012}.
The images are then re-binned into 36\arcsec\ pixels, chosen to be about the same size as the 500\micron\ beam
so that each pixel is approximately independent from each other. For each band we subtract a background value for the whole map, estimated 
from regions around the galaxy.
The uncertainties on the flux in each pixel are found by measuring the standard deviation of the pixels
in these background regions and adding this in quadrature with the calibration uncertainty. 
The flux errors in the majority of pixels are dominated by the calibration uncertainty which we take as
7\% for \Spitzer\ 70\micron\ \citep{Gordon2007}, 10\% for PACS \citep[see][]{Fritz2011} and 7\% 
for SPIRE (see Section \ref{sect:SEDfit} for more details).

\subsection{SED fitting}
\label{sect:SEDfit}
For each pixel we fit
the far-infrared--sub-millimetre SED with a one-temperature modified black-body model
described by:
\begin{equation}
S_{\nu} = {{\kappa_{\nu}M_dB({\nu},T_d)}\over{D^2}}
\label{equ:SED}
\end{equation}
where $M_d$ is the dust mass with dust temperature $T_d$,
$B(\nu,T_d)$ is the Planck function and $D$ is the distance to the
galaxy. $\kappa_{\nu}$ is the dust absorption
coefficient, described by a power law with dust emissivity index
$\beta$ such that $\kappa_{\nu} \propto \nu^{\beta}$. We assume a typical value 
for the ISM of $\kappa_{350\mu \rm m} = 0.192\,\rm m^2\,kg^{-1}$ \citep{Draine2003}. While
the absolute value of $\kappa_{\nu}$ is uncertain, its value will not affect trends with other parameters as it is a fixed
scaling constant. The distance to Andromeda was taken in this work to be $D=0.785\,\rm Mpc$ \citep{McConnachie2005}.

We initially used a fixed value of $\beta$ across the whole galaxy, but we found
that with a fixed value it was impossible to adequately fit the SEDs; we therefore let $\beta$ be a free parameter. To ensure
the simplex fitting routine did not get stuck in a local minimum, we ran the SED-fitter in two ways: 
first with all three parameters (M$_d$, T$_d$, $\beta$) free to vary; second by fixing the value of $\beta$
while allowing M$_d$ and T$_d$ to vary
and repeating the process for all values of $\beta$ in the range 0.20 to 5.90 in 0.01 intervals, selecting 
the result with the lowest $\chi^{2}$. Both methods gave consistent
results but we created the final maps of the dust properties by choosing the result with the lowest $\chi^{2}$ for each pixel.
  
The SPIRE calibration has an overall systematic uncertainty of 5\% due to the uncertainty in the prime calibrator Neptune, and a statistical uncertainty of 2\%
determined from instrumental reproducibility; 
the SPIRE Observer's Manual recommends linearly adding these to give an overall uncertainty of 7\%. To implement this in practice 
in our fitting algorithm we increased
the uncorrelated uncertainty to give the same overall uncertainty when the errors are added in quadrature. This is 
implemented in the SED fitting by using the full covariance matrix in the $\chi^{2}$ calculation.

We apply our own color correction to the \Hersc\ maps by removing the \Hersc\ pipeline `K4' parameter and then convolving the SED
with the filter transmission in the fitting process (for SPIRE the filters appropriate for extended sources are used).
This method takes full account of all the wavelength-dependant effects associated with PACS and SPIRE. 
In previous work \citep{Smith2010} we found
that there is a significant contribution from a warmer component of dust at wavelengths $\leq$\,70\micron\
and so the \Spitzer\ flux at 70\micron\ is treated as an upper limit in the fitting (i.e, if the flux value is higher than the model it does not contribute to $\chi^{2}$). 
The omission of the warm component
in the fitting process has a negligible effect on the derived dust mass as the cold component
dominates the total dust mass.
\citet{Bendo2012} suggest that a warmer
component could influence the dust emission up to wavelengths of 160\micron; to test this we repeated the SED-fitting by treating 
the \Spitzer\ and PACS fluxes as upper limits and found it made a negligible difference to our results.

To estimate the uncertainties on the results of our fits we use a monte-carlo technique. For each pixel we create a set of 1000 artificial SEDs, created by taking the
original flux densities and adding a random value selected from a normal distribution with a mean of zero and a standard deviation equal to the uncertainty in the measured flux
(the correlations between the calibration uncertainties for SPIRE are also included). 
We estimate the error in the derived parameters for each pixel from the 1000 fits. 
We find that for each pixel there is an uncertainty of 20\% in our estimate of the surface density of the
dust, of $\pm$1.4\,K in our estimate of the dust temperature, and $\pm$0.31 in our estimate of $\beta$.

\subsection{Results of the Fits}

\label{sec:fit-analysis}

\begin{figure}
  \centering
  \includegraphics[trim=1mm 1mm 5mm 4mm,clip=true,width=0.49\textwidth]{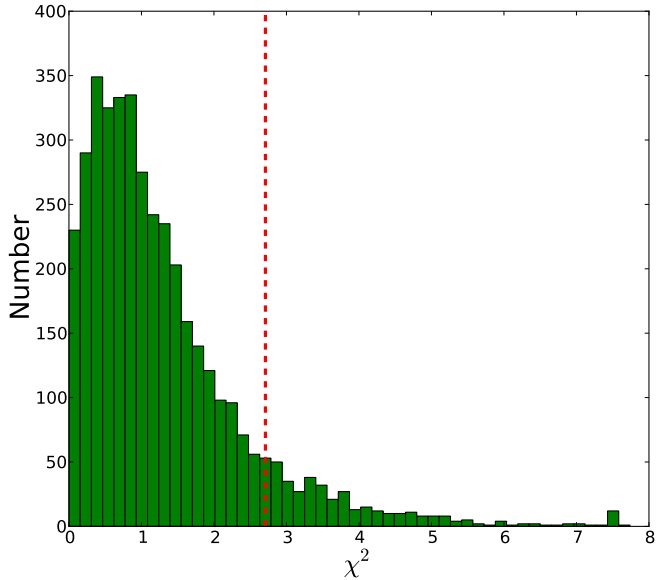}
  \caption{Distribution of $\chi^{2}$ values from pixels fitted with the modified-blackbody model. The red line represents the 10\% significance value for one degree of freedom.}
  \label{fig:chisq}
\end{figure}

In producing the final dust mass, temperature and $\beta$ images in this work, we only used pixels where the fluxes in all six bands 
(five \Hersc\ \& MIPS 70\micron) have a signal-to-noise greater than 5$\sigma$. 
While this potentially causes us to miss the very coldest dust due to weak emission in the shortest wavelength bands,
we choose it as a conservative approach to ensure we have accurate estimates of temperature.
In practice it is the 100\micron\ map, which has the lowest sensitivity, which limits the number of pixels in our selection.
Despite this very conservative selection, there are still $\sim 4000$ pixels
in our resultant maps. To see if our model is a statistically reasonable fit to the data, we created a
histogram of the $\chi^{2}$ values for all pixels (Figure \ref{fig:chisq}). As the 70\micron\ flux is used as an upper limit (see Section \ref{sect:SEDfit}) and is usually
higher than the best-fit model it does not usually contribute to $\chi^{2}$ and therefore we only have 1 degree of freedom ($5_{\rm data\ points} - 3_{\rm parameters} - 1$).
The 10\% significance level for $\chi^{2}$ occurs at 2.71 which is shown by the red line in Figure \ref{fig:chisq}. We find 9.8\% of our fits
have $\chi^{2}$ above this level showing our model is an adequate representation of the data. We have also checked to see if our radial gradients in temperature,
$\beta$, and dust surface density are affected by lowering the criteria to include fluxes greater than 3$\sigma$ and find no significant changes.

Recent results from the Key Insight on Nearby Galaxies: A Far-Infrared Survey with \Hersc\ (KINGFISH) presented in \citet{Dale2012} have suggested that the
one-temperature modified black-body model underestimates the dust mass compared to the \citet{DraineLi2007} prescription. They attribute this difference
to the contribution of warm dust emitting at shorter wavelengths. For our analysis this does not appear to be the case. First when we set the PACS fluxes
($\leq$\,160\micron) as upper limits there is no significant change in our results. Second, if multiple temperature components were present this would bias 
our $\beta$ values to lower values \citep[since each temperature component peaks at a different wavelength,][]{Shetty2009}, 
but we mostly find higher $\beta$ values 
than expected (see Section \ref{sec:res}). Third the $\chi^{2}$ analysis suggests
the model is consistent with the data. In addition, \citet{Mentuch2012}, using a similar analysis, have found in M51 that the dust mass distribution is similar when using 
the \citet{DraineLi2007} or the single modified blackbody prescription.

\begin{figure}
  \centering
  \includegraphics[trim=4mm 0mm 18mm 18mm, clip=true, width=0.49\textwidth]{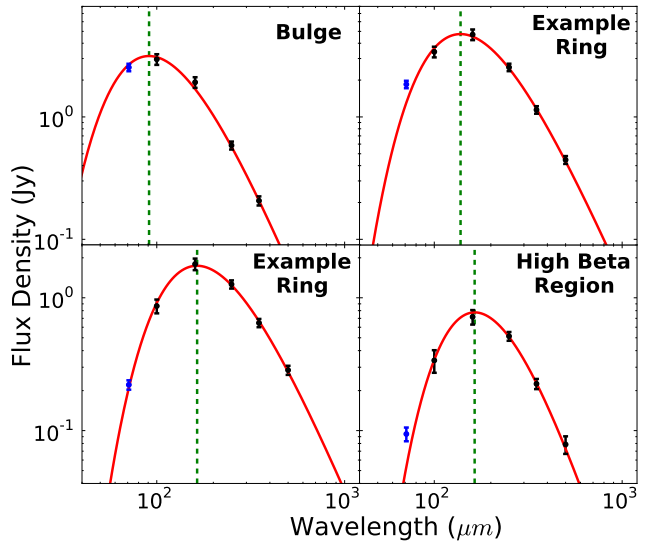}
  \caption{Examples of SED fits for pixels in different regions. The 70\micron\ point (blue) is used as an upper limit 
           and the peak of distribution is shown by the dashed green line.}
  \label{fig:exampleSEDs}
\end{figure}

We applied the same fitting technique as outlined in Section \ref{sect:SEDfit} to the global flux densities, measured in an elliptical aperture
with a semi-major axis of 2.0$^\circ$ and semi-minor axis of 0.73$^\circ$.
This produces a total dust mass of $10^{7.86\pm 0.09} M_{\odot}$ with a dust temperature of $16.1\pm 1.1$\,K and $\beta = 1.9\pm 0.3$. 
The total dust mass from summing all the pixels in our pixel-by-pixel analysis gives a value $10^{7.46} M_{\odot}$, a factor of $\sim$2.5 lower. This
is expected as the fraction of 500\micron\ flux in the pixels which satisfy our signal to noise criterion is approximately half the global value.
Combining our global dust mass with gas measurements, we find a global gas-to-dust mass ratio of 72.
The global temperature is consistent
with these for other spiral galaxies obtained using similar methods \citep[e.g.,][]{Davies2011}. 
The pixel-by-pixel analysis shows a large range of temperatures and $\beta$ values as discussed in detail in Section \ref{sec:res}. Examples of fits for individual pixels are shown
in Figure \ref{fig:exampleSEDs}.

\subsection{Simulations}
\label{sec:simulation}

To help understand the significance of our results and any potential biases or degeneracies in the parameters, we ran
a Monte-Carlo simulation. Assuming the dust emits as a single-component modified blackbody, we generated synthetic flux values
for a range of temperatures and $\beta$ values, with the same wavelengths as our real data. 
Noise was then added to the simulated fluxes with a value equal to the errors in the real fluxes
(excluding the calibration errors) for 2000 repetitions per T, $\beta$ combination. 
The calibration error was not included as it would systematically shift the fluxes for all pixels. 
We chose an input mass surface density
of 0.5\,$M_{\odot}\rm\ pc^{-2}$ as this roughly corresponds to the values found in the 10\,kpc ring. 

\begin{figure}
  \centering
  \includegraphics[trim=2mm 3mm 0mm 3mm,clip=true,width=0.49\textwidth]{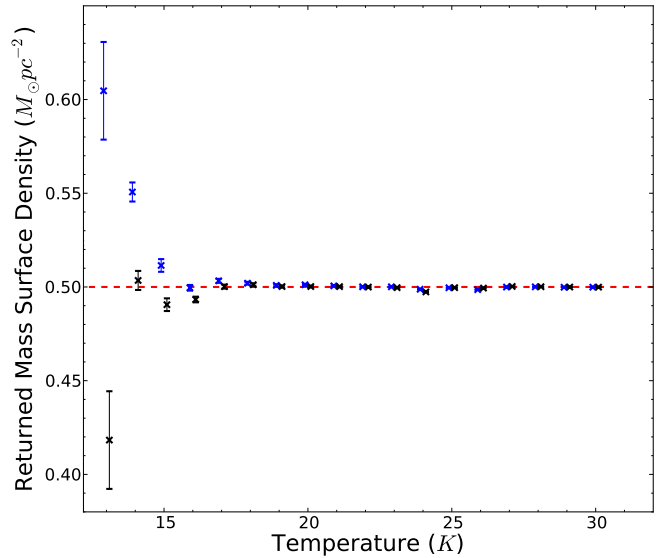}
  \caption{The range of mass surface densities returned from the SED fitting technique 
           versus the input temperature for the simulated data in Section \ref{sec:simulation} with $\beta = 2$ modified-blackbody. 
           The mean of the returned masses is shown by the blue points and the median
           is shown by the black points. The input mass surface density of 0.50 $M_{\odot}\rm\ pc^{-2}$ is shown by the red dashed line. To avoid overlapping data points the blue
           have been shifted by -0.1\,K and the black by +0.1\,K.}
  \label{fig:sim-mass}
\end{figure}

The quantity that is most important for our work is the dust mass, which from Equation \ref{equ:SED} is just a multiplicative term. 
Figure \ref{fig:sim-mass} shows the mean and median mass returned for the 2000 repetitions as the input temperature is varied between 
12 and 30\,K for a $\beta$ of 2; 
the error bars show the error on the mean. 
Between 15 and 30\,K, the mean and median dust mass returned matches within the errors the input dust surface density of 0.50 $M_{\odot}\rm \ pc^{-2}$.

At dust temperatures of 15\,K and below, there are large errors on the returned mass, which is due to the fluxes at the PACS wavelengths not reaching the required
sensitivity to be included in the fit. 
For the actual data, we only included pixels in which there is at least a 5$\sigma$ detection in all bands, which will avoid the wildly incorrect
estimates of the dust mass seen in the simulation. To fully estimate the dust mass and temperature of very cold dust (T $<$ 15\,K), 
we need flux measurements at $>$500\micron. Nevertheless, the lack of an excess at 500\micron\ (Section \ref{sec:excess}) 
is circumstantial evidence that Andromeda does not contain very cold dust. 
We investigated if the results in Figure \ref{fig:sim-mass} were different for a different choice of $\beta$ but found no systematic differences. 

By plotting the difference in the resultant temperature in each pixel compared to the input model 
temperature, we find that between 15\,K and 25\,K, the temperature uncertainty is $\sim$0.6\,K.
Above 25\,K the uncertainty increases, although
in the simulation we did not include a 70\micron\ point which would likely provide a constraint to our fits if the dust temperature was $>$25\,K. 
The simulation suggests that for input $\beta$ between 1.5 and 2.4, the uncertainty in the returned value of $\beta$ for each pixel is $\sim$0.1. 
As expected these uncertainties are lower than
returned by the monte-carlo technique in Section \ref{sect:SEDfit} as we have not included a calibration uncertainty.

\begin{figure}
  \centering
  \includegraphics[trim=2mm 1mm 23mm 22mm,clip=true,width=0.49\textwidth]{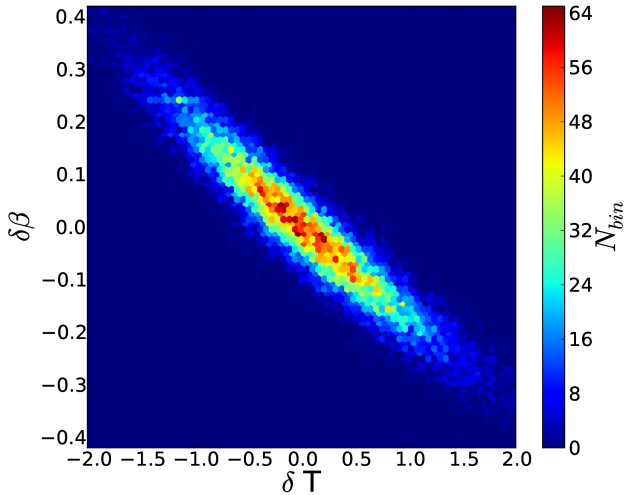}
  \caption{A density plot showing the correlated uncertainties between $\beta$ and temperature. The uncertainties
           are taken from the simulated data with $T=17$\,K and $\beta=1.8$. A clear anti-correlation is observed with the
           distribution centered on the correct values.
           To fully populate this graph we increased the simulation to 20,000 repetitions for this $\beta$, T combination.}
  \label{fig:sim-delt-delt}
\end{figure}

A degeneracy is known to exist between temperature and $\beta$; Figure \ref{fig:sim-delt-delt} shows that if there is an error in one parameter 
this is anti-correlated with the error in the other parameter. As the distribution is centered on the origin,
there is no systematic offset in the returned value of T or $\beta$. 
As there is no systematic offset, the error in the mean values over many pixels will 
therefore be much smaller than the error for a single pixel.

This simulation is based on the dust emission arising from a one-component modified blackbody. Fitting a one-component modified blackbody to pixels for which there are
multiple temperature components would produce a bias towards smaller values of $\beta$ \citep{Shetty2009}, although we would hope to detect this by finding 
high $\chi^{2}$ values. 
In Section \ref{sec:beta-T} we show that different regions of M31 have different $\beta - T$ relations and discuss why this is unlikely to be due to multiple
temperature components. 

To summarise, while there is a $\beta$--T degeneracy from the fitting algorithm this does not create any systematic offsets in the value returned.
If the dust temperature falls below 15\,K we are unable to constrain the SED due to the lack of data points beyond 500\micron.

\begin{figure*}
  \centering
  \includegraphics[trim=1.0mm 2.0mm 0.0mm 2.0mm,clip=true,width=1.0\textwidth]{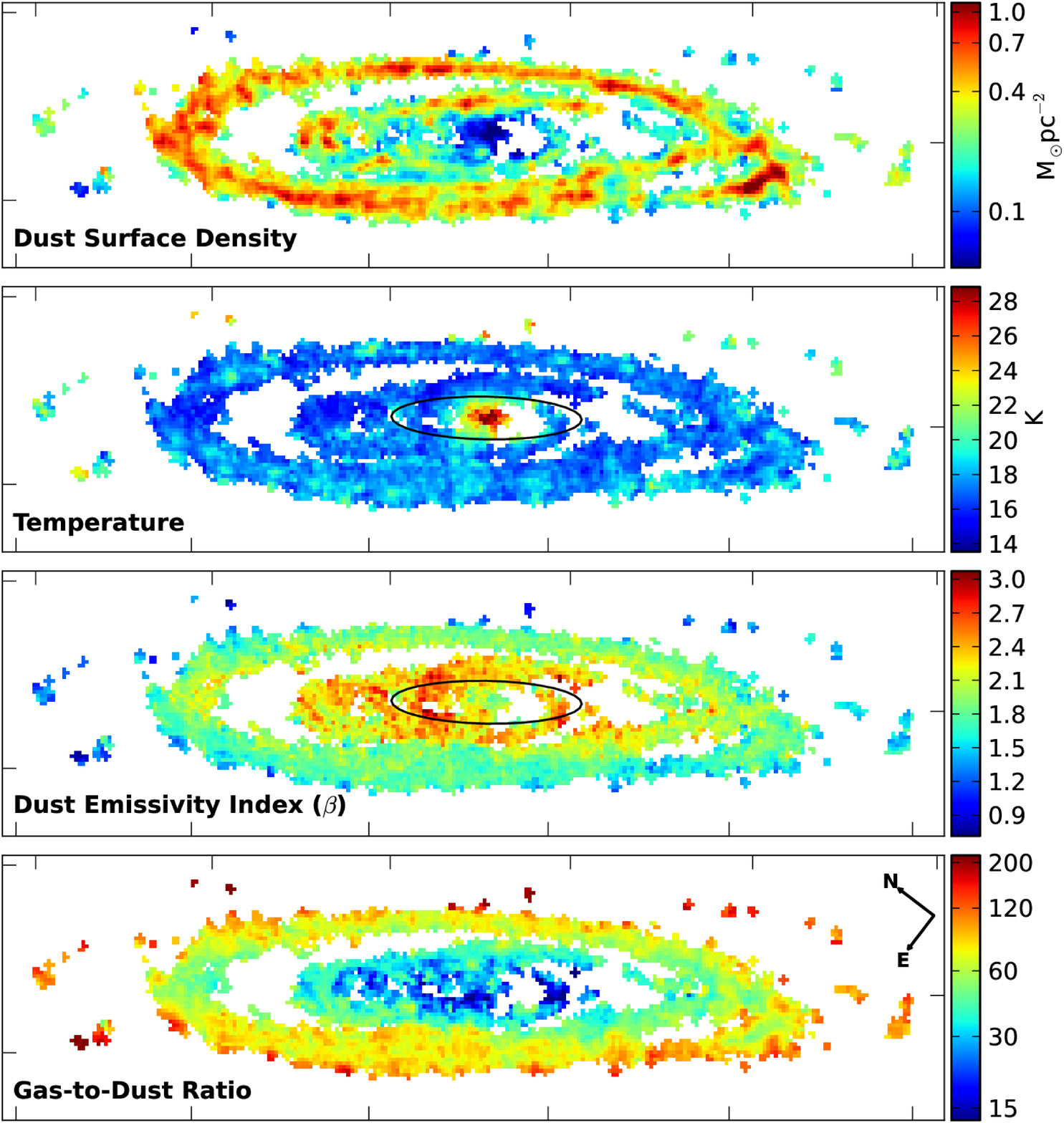}
  \caption{The distribution of dust surface density, temperature and $\beta$ obtained from the SED fitting technique and 
           the distribution of the gas-to-dust ratio. The temperature and $\beta$ images include a black ellipse showing a radius of 3.1\,kpc. 
           The ticks are plotted at 30\arcmin\ intervals.}
  \label{fig:SED-maps}
\end{figure*}

\section{Results}
\label{sec:res}

\subsection{Spatial Distribution of Dust Mass, Temperature and Emissivity Index}

\label{sec:spatial-dist}

By fitting SEDs to every pixel, we have created maps of dust surface density, temperature, $\beta$ and gas-to-dust ratio which are shown in
Figure \ref{fig:SED-maps} (for details on how the gas surface density is calculated see Section \ref{sec:radGD}). The dust surface density distribution, un-surprisingly, is more 
similar to the maps of gas and star-formation rate 
than to the 3.6\micron\ image (See Figure \ref{fig:her-maps} \& \ref{fig:oth-maps}), which traces the stellar mass distribution.

\begin{figure}
  \centering
  \includegraphics[trim=0mm 3mm 0mm 3mm,clip=true,width=0.49\textwidth]{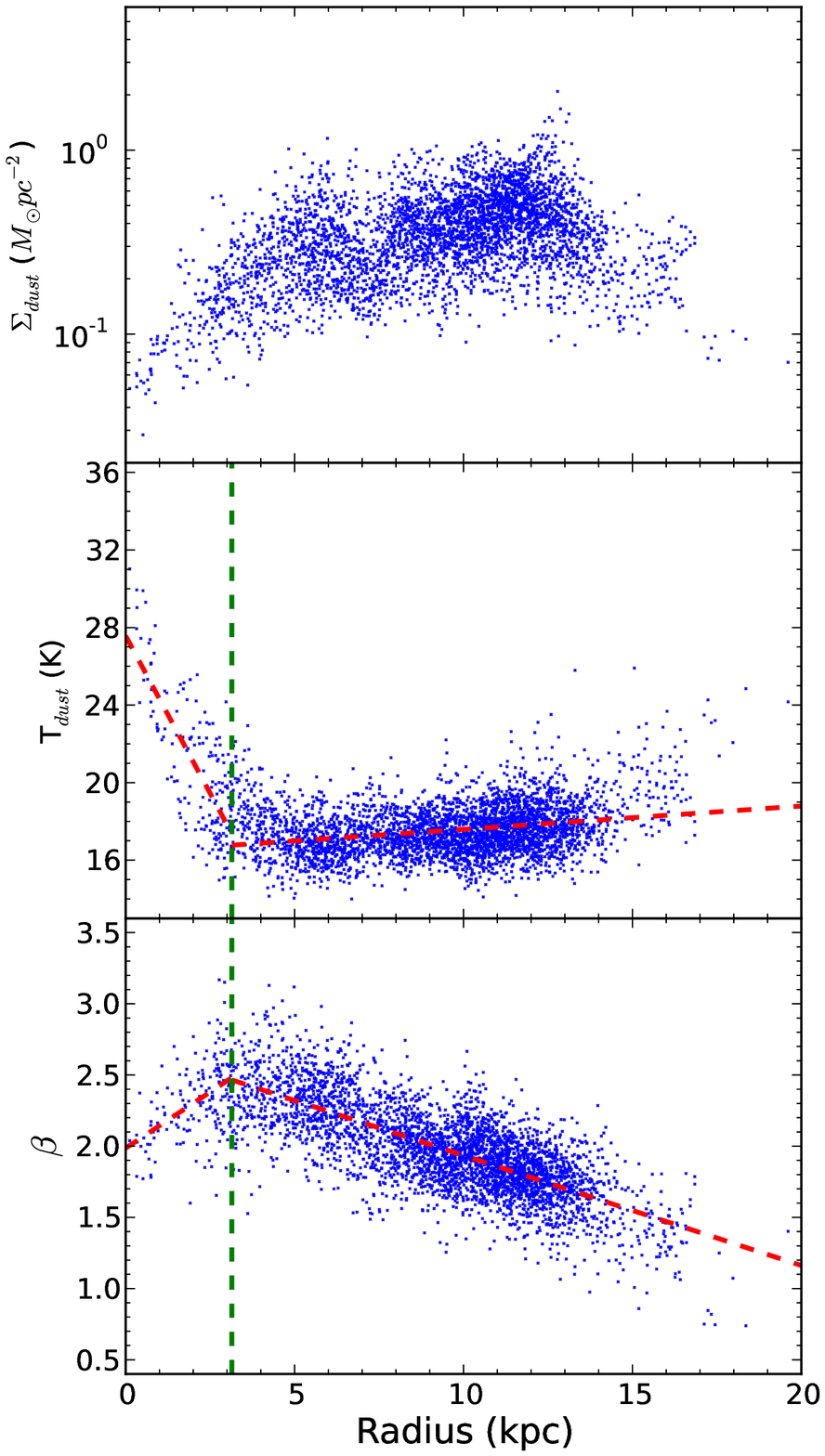}
  \caption{Results from the SED fits for each pixel plotted versus radius. The dashed red lines represent the best-fit linear model (see Section \ref{sec:heating}), the
           dashed green line represents the transition radius (3.1\,kpc) found when fitting the $\beta$ results.}
  \label{fig:sed-rad}
\end{figure}

The $\beta$ and temperature maps show significant radial variations. Figure \ref{fig:sed-rad} shows how the dust column density, temperature and $\beta$ vary with radius
(the physical radius is calculated assuming an inclination of 77$^{\circ}$ and PA of 38$^{\circ}$, \citealt{Fritz2011}). 
In the star-forming ring at 10\,kpc, $\beta$ has an average value of $\sim$1.8 but increases with 
decreasing radius reaching a maximum of $\sim$2.5 at a radius of $\sim$3\,kpc. This is higher than found in global studies of galaxies \citep{Planck2011, Davies2011, Dunne2011}. 
However similarly high values have 
been reported recently in \citet{Foyle2012} and \citet{Bracco2011} for dust within galaxies. 
The value for the ring agrees well with early results from \textit{Planck} 
\citep{Planck2011} for dust in the galactic disk and the solar neighbourhood. The 10\,kpc ring
has an average dust surface density of $\sim$0.6\,$M_{\odot} \rm\ pc^{-2}$ with a dust temperature of 18\,K.
Towards the very center of the galaxy the dust column density decreases to $\sim$0.04\,$M_{\odot} \rm\ pc^{-2}$, 
$\beta$ values fall to $\sim$1.9 and the dust temperature increases to $\sim$30\,K. 

\begin{figure}
  \centering
  \includegraphics[trim=0mm 3mm 0mm 3mm,clip=true,width=0.49\textwidth]{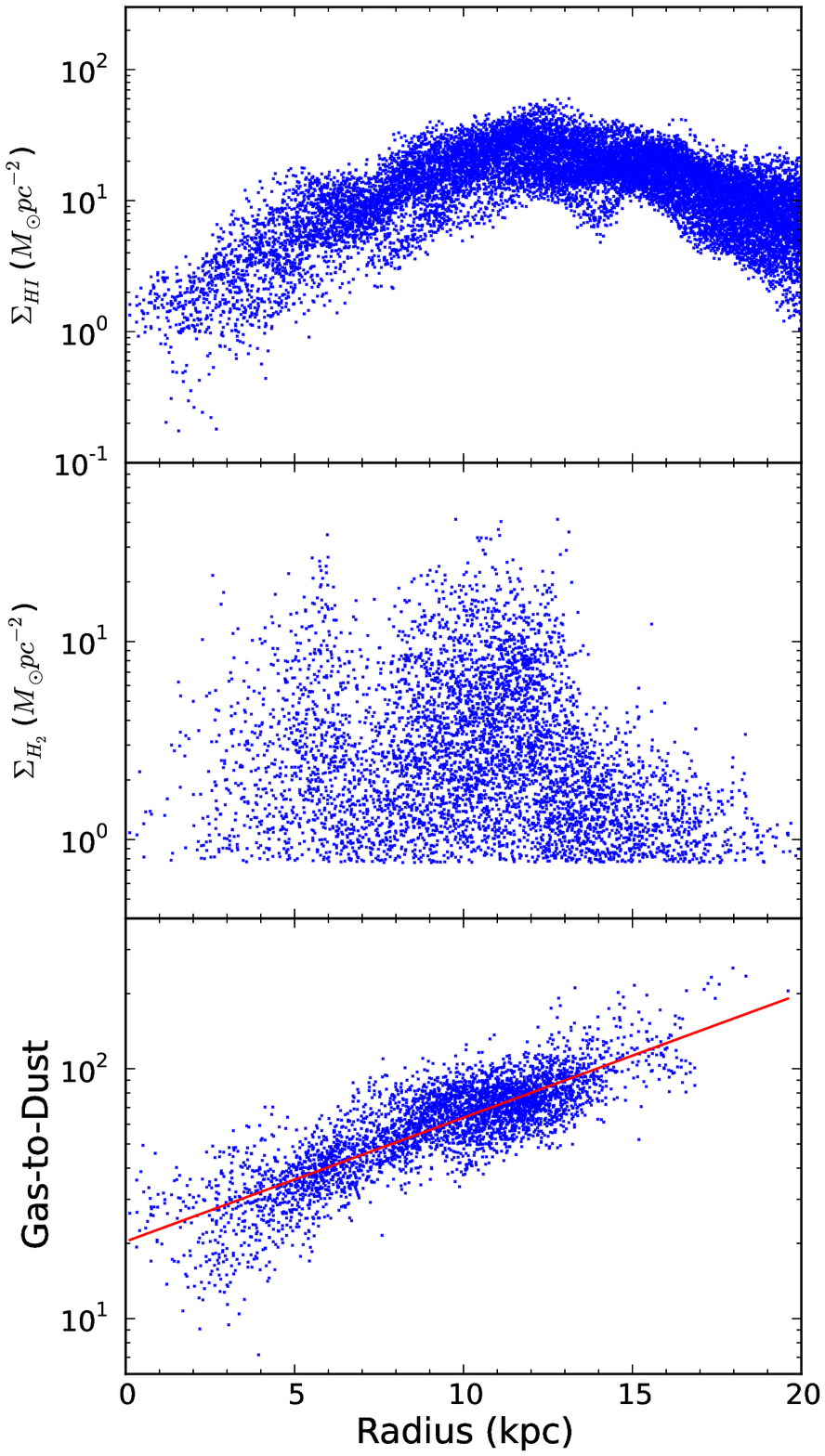}
  \caption{The distribution of atomic gas, molecular gas and gas-to-dust ratio versus radius. The two 
           gas maps are plotted for all pixels $>3\sigma$. The solid red line
           represents the best-fit exponential profile to the gas-to-dust ratio.}
  \label{fig:gas-rad}
\end{figure}

In Figure \ref{fig:sed-rad} we see a clear break in the radial variation for both temperature and $\beta$ at a 
radius $\sim$3\,kpc. We fit a model with two straight lines and a transition radius (using a simplex routine) to the
$\beta$ results. The same method is used with the temperature values but we set the transition radius to the value obtained from the fit to the $\beta$ values,
which occurs at 3.1\,kpc (shown by the dashed green line in Figure \ref{fig:sed-rad} or black ellipse in Figure \ref{fig:SED-maps}). At radii smaller than the transition radius, the temperature 
decreases with radius from $\sim$30\,K in the center to $\sim$17\,K, with an associated increase in $\beta$ from $\sim$1.8 to $\sim$2.5.
At radii larger than the transition radius the dust temperature slowly increases with radius while $\beta$ decreases with radius to $\sim$1.7 at 13\,kpc. 
The best-fit relationships between $T_d$, $\beta$ and $\rm {\it R}$ are shown in Figure \ref{fig:sed-rad} and listed below:
\begin{align}
\label{equ:profiles1}
\ \ \, \beta & = ~~0.15{\rm \it R} ({\rm kpc}) + 1.98  ~~~~~~~~~~{\rm \it R}\, < \,3.1\,{\rm kpc} \\
\label{equ:profiles2}
\ \ \, & =  -0.08 {\rm \it R} ({\rm kpc}) + 2.70 ~~~~~~~~~~3.1\le {\rm \it R}\, < \,20\,{\rm kpc} 
\end{align}
\begin{align}
\label{equ:profiles3}
 T_d & = -3.24 {\rm \it R} ({\rm kpc}) + 27.56 ~~~~~~~~{\rm \it R}\, < \,3.1\,{\rm kpc}\\
 & = ~~\,0.12{\rm \it R} ({\rm kpc}) +16.40~~~~~~~~3.1\le {\rm \it R}\, < \,20\,{\rm kpc} 
\label{equ:profiles4}
\end{align}
We discuss the cause of the temperature and $\beta$ variations in Section \ref{sec:disc}.
For Sections \ref{sec:heating} \& \ref{sec:beta-T} we consider the inner (${\rm \it R}$\,$<$\,3.1\,$\rm kpc$) and outer regions separately.

\subsection{Radial Distribution of Gas and Dust}
\label{sec:radGD}

In Figure \ref{fig:gas-rad} the radial variations of the atomic gas, the molecular gas and 
the gas-to-dust ratio are shown. We assume an X-factor of $1.9\times 10^{20}\rm\,cm^{-2}\,[K\, kms^{-1}]^{-1}$
\citep{Strong1996} to convert a CO line flux to a H$_2$ column density, although this value is notoriously uncertain and has been found to vary with metallicity 
\citep{Strong2004, Israel2005}. For our analysis of M31, the choice of X-factor makes very little difference as the total gas is dominated by the atomic phase. Out of the 3974 pixels 
plotted in the gas-to-dust figure, only 101 have a molecular fraction of $>$50\% and globally the molecular gas only constitutes $\sim$7\% of the atmoic gas \citep{Nieten2006}.
To estimate the total gas surface density we include the contribution
of the atomic or molecular gas in each pixel if the value is greater than 3$\sigma$ in their respective maps. Only 86 of our 5$\sigma$ dust pixels are not covered by the CO($J$=1-0) observations; 
these pixels are in the outskirts of the galaxy and we assume the contribution of molecular gas is negligible.
We find a tight relation between the gas-to-dust ratio and radius
(Spearman Rank Coefficient of 0.80) as shown in Figure \ref{fig:gas-rad}, which is described by an exponential profile, shown by the red line.

The gas-to-dust ratio increases exponentially from low values of $\sim$20 in the center of the galaxy
to $\sim$110 at 15\,kpc typical of the Milky Way in the local environment \citep[see][and references therein]{Devereux1990}. Note that the values are not corrected for helium in the ISM. Metallicity
gradients have been estimated from oxygen line ratios [OIII]/H$\beta$, [OII]/[OIII] and R$_{23}$ by \citet{Galarza1999}. They infer
a radially decreasing metallicity gradient of $-0.06 \pm 0.03 \rm{\ dex\ kpc^{-1}}$ from the R$_{23}$ parameter. 
\citet{Trundle2002} calculate oxygen abundance gradients based on a set of 11 H{\sc ii} regions from \citet{Blair1982} and find
gradients in the range of -0.027 to -0.013\,$\rm dex\ kpc^{-1}$ depending on the calibration used.
If a constant fraction of the metals in the ISM is incorporated into dust grains \citep{Edmunds2001}, one would expect the gas-to-dust gradient to be $-1\times$ metallicity gradient. 
We find a gas-to-dust gradient of $0.0496 \pm 0.0005\rm{\ dex\ kpc^{-1}}$ consistent within the uncertainties to the gradient measured by \citet{Galarza1999}
(if the H{\sc i} map corrected for self-opacity is used the gradient slightly increases to $0.0566 \pm 0.0007\rm{\ dex\ kpc^{-1}}$).
This supports the claim that gas can be well traced by dust at a constant metallicity. 
To see if the uncertainty in the choice of X-factor could affect this result, we carried out the same procedure but limited the analysis to pixels where the molecular hydrogen contribution is
less than 10\% of the total gas mass. This produced only a small change in the gas-to-dust gradient to $0.0550 \pm 0.0007\rm{\ dex\ kpc^{-1}}$ which is still consistent with the metallicity gradient.

\subsection{500\micron\ Excess}

\label{sec:excess}

\begin{figure}
  \centering
  \includegraphics[trim=0.0mm 0.0mm 0.0mm 0.0mm,clip=true,width=0.5\textwidth]{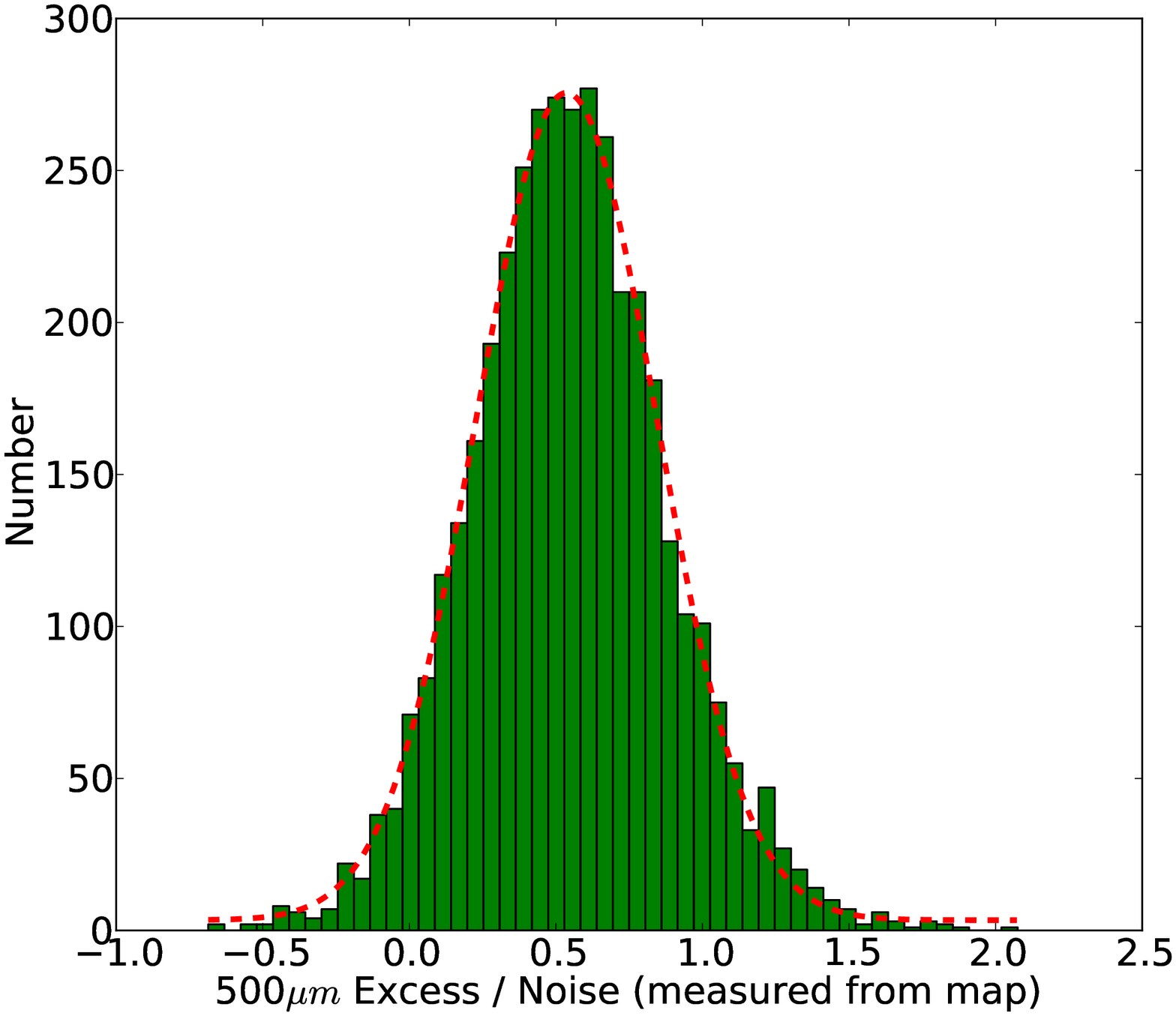}
  \caption{The value of the 500\micron\ excess, defined as the 500\micron\ - flux from our blackbody model, divided by the noise. The
           red line is a Gaussian fit to the histogram.}
  \label{fig:excess}
\end{figure}

Searches for a long-wavelength sub-mm excess (i.e. $>$500\micron) has been made in both the Milky Way \citep{Paradis2012} and for nearby galaxies. A sub-mm excess is important as it 
could suggest very cold dust is present ($<$15\,K) which would dominate the dust mass in galaxies. Other possible explanations of an excess
include variations in the dust emissivity index with wavelength \citep{Wright1991, Reach1995, Paradis2012}, different dust populations or contamination from a synchrotron radio source.
Sub-mm excesses have been reported from observations of low-metallicity
dwarf galaxies \citep[see][]{O'Halloran2010, Grossi2010, Dale2012} and spiral galaxies \citep[see][]{Zhu2009,Bendo2006}.
Most detections have been made by combining FIR data with ground-based data at 850\micron\ or 1\,mm data rather than from
data only at $\le$500\micron.

We searched for a sub-millimetre excess in M31 by comparing the 500\micron\ flux to our best-fit models.
A 500\micron\ excess is defined to be any observed 500\micron\ flux that is greater than the expected flux at 500\micron\ from our
modified blackbody fit to the data.
Figure \ref{fig:excess} is a histogram of the ratio of the excess at 500\micron\ and the noise on the 500\micron\ map.
The distribution of the excess is consistent with a Gaussian function with a mean of 0.54$\sigma$ and standard deviation of 0.31$\sigma$. The fact the
distribution of the histogram is a Gaussian suggests that what we are seeing is random noise with a fixed offset (not centered on zero). 
Two non-astronomical sources could explain a fixed offset, either
an incorrect background subtraction or a calibration error. The background correction applied is quite small (0.2$\sigma$),
and thus an error in this is unlikely to be the entire explanation of the offset. 
The distribution of the excess is consistent with the 2\% random SPIRE calibration uncertainty, 
so both factors together could explain the small offset. If the excess is generated from dust with a 10\,K temperature and $\beta$ of 2,
the dust mass is $\sim$10$^{6.58}$\,$M_{\odot}$ in our $>$5$\sigma$ pixels, which corresponds to only 13\% of the mass of the warmer dust.
If we used a model containing dust at more then one temperature we would not get a useful constraint on the colder dust component 
without additional data at longer wavelengths. In particular observations at $\sim$850\micron\ (e.g. with SCUBA2) would be useful to determine if any cold dust present.

\section{Discussion}
\label{sec:disc}

\subsection{Heating Mechanisms and Dust Distribution}
\label{sec:heating}

Recent studies by \citet{Bendo2010b, Bendo2012} and \citet{Boquien2011} have used \Hersc\ colors to confirm previous works \citep[e.g.,][]{Persson1987, Walterbos1996} 
that emission from dust in nearby galaxies at wavelengths longer than 160\micron\ is mostly from dust heated by the general stellar population. 
These authors conclude that at wavelengths shorter than 160\micron, the emission tends to be dominated from warmer dust heated by newly formed stars. 
\citet{Montalto2009} in a study of M31 using \Spitzer MIR/FIR, UV and optical data, also concluded that the dust is mostly heated by an old stellar population (a few Gyr old).
To investigate the relation between our derived dust properties and the SFR and the general stellar population,
we have used the 3.6\micron\ map presented in \citet{Barmby2006} to trace the general stellar population 
(rather than the luminous newly formed stars that dominate the UV) and 
a map of star-formation rate (SFR). The SFR map is created from far-UV and 24\micron\ images which have been corrected for the contribution of 
the general stellar population to the emission at these wavelengths (for details see G. P. Ford et al. in preparation). 
These maps were all convolved to the same resolution and binned to the same pixel size as all the other maps in our analysis.

\begin{figure*}
  \centering
  \includegraphics[trim=5.0mm 3.0mm 0.0mm 0.0mm,clip=true,width=1.0\textwidth]{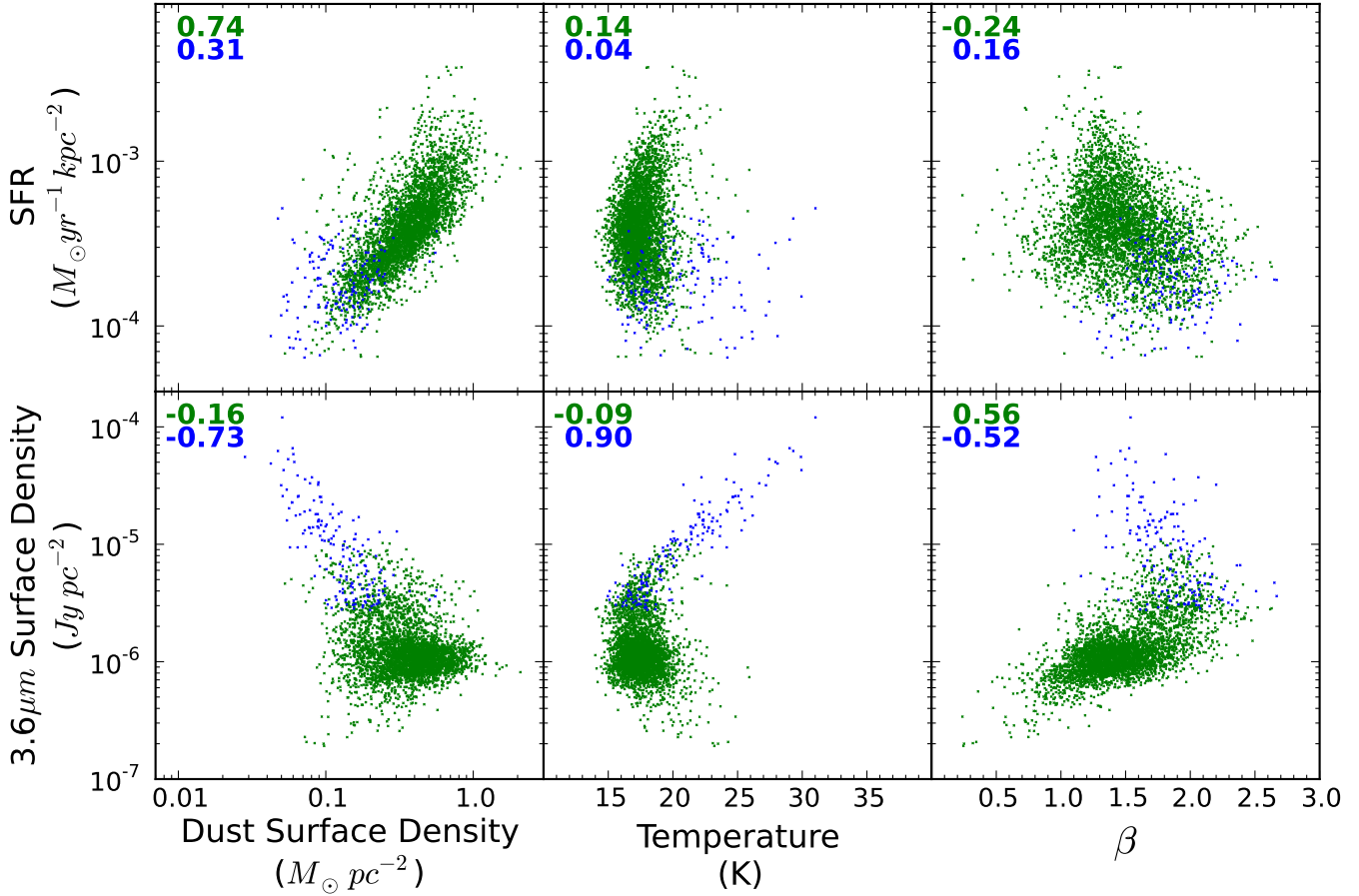}
  \caption{Scatter plots showing correlations between dust properties and 3.6\micron\ flux or SFR. The blue points are results at radii $<$3.1\,kpc and 
           the green data points are results at radii $>$3.1\,kpc. The Spearman rank-order coefficients for both sets of points are shown in the top-left corner
           of each plot.}
  \label{fig:heating}
\end{figure*}

We have plotted the fluxes from these maps against the results of our SED-fits (see Figure \ref{fig:heating}). 
In Section \ref{sec:spatial-dist} we showed there is a clear break in the dust properties at a radius of 3.1\,kpc. 
Therefore in Figure \ref{fig:heating} the pixels at a radius less than 3.1\,kpc are shown in blue and those at a radius 
above 3.1\,kpc in green. For each graph (and both sets of pixels) the Spearman rank coefficient is computed 
and an estimate made of its significance (see Table\,\ref{table:spearman}). With such a large number of pixels, all but
one of our correlations are formally significant. In the discussion below we have concentrated on the strongest correlations
as measured by the Spearman correlation coefficient.

There are strong correlations in Figure \ref{fig:heating} between $\beta$ and 3.6\micron\ emission. We believe these are 
most likely to be caused
by radial variations in $\beta$ seen in Figure \ref{fig:sed-rad} and the decrease in 3.6\micron\ brightness with radius.
We discuss the possible cause of the radial variation in $\beta$ in Section \ref{sec:beta-T}.

We find a strong correlation between dust surface-density and the SFR in the outer regions,
but not with the surface density of total stars traced by the 3.6\micron\ emission. This correlation is expected as stars are formed in clouds of gas and dust. 
In the inner region there is an anti-correlation between the dust surface density and the 3.6\micron\ flux, 
which seem most likely to be explained by both quantities varying with radius: the 3.6\micron\ emission from the bulge increasing 
towards the center and the dust surface density decreasing towards the center.

\begin{deluxetable}{ccccc}
\tabletypesize{\scriptsize}
\tablecaption{Spearman Correlation Coefficients for Properties of M31\label{tab:spearman}}
\tablewidth{0pt}
\tablehead{
\colhead{Property A} & \colhead{Property B} & \colhead{Region} & \colhead{Spearman} &\colhead{P-Value}\\
\colhead{} & \colhead{} & \colhead{} & \colhead{Coefficient} & \colhead{}}
\startdata
\multirow{4}{1.0cm}{\centering{Dust Surface Density}} & \multirow{2}{*}{3.6\micron\ flux} & Inner & -0.73 & $1.52\times 10^{-28\ }$\\
 & & Outer & -0.16 & $1.53\times 10^{-22\ }$\\
 & \multirow{2}{*}{SFR} & Inner & \ 0.31 & $6.38\times 10^{-5\ \ }$\\
 & & Outer & \ 0.74 & 0.00\\
\cmidrule(rl){1-5}
\multirow{4}{*}{Temperature} & \multirow{2}{*}{3.6\micron\ flux} & Inner & \ 0.90 & $6.17\times 10^{-59\ }$\\
 & & Outer & -0.09 & $7.46\times 10^{-9\ \ }$\\
 & \multirow{2}{*}{SFR} & Inner & \ 0.04 & $5.97\times 10^{-1\ \ }$\\
 & & Outer & \ 0.14 & $3.70\times 10^{-19\ }$\\
\cmidrule(rl){1-5}
\multirow{4}{*}{$\beta$} & \multirow{2}{*}{3.6\micron\ flux} & Inner & -0.52 & $8.78\times 10^{-13\ }$\\
 & & Outer & \ 0.56 & $8.99\times 10^{-311}$\\
 & \multirow{2}{*}{SFR} & Inner & \ 0.16 & $3.58\times 10^{-2\ \ }$\\
 & & Outer & -0.24 & $2.85\times 10^{-52\ }$
\enddata
\label{table:spearman}
\tablecomments{The Spearman rank-order correlation coefficient and P-value (for the the null hypothesis that the two data sets are uncorrelated) 
for the scatter plots shown in Figure \ref{fig:heating} (values were calculated using the scipy.stats package and checked with IDL r\_correlate routine).
The inner and outer region contain 164 and 3810 pixels, respectively.}
\end{deluxetable}

\begin{figure}
  \centering
  \includegraphics[trim=0.0mm 3.0mm 0.0mm 0.0mm, clip=true,width=0.5\textwidth]{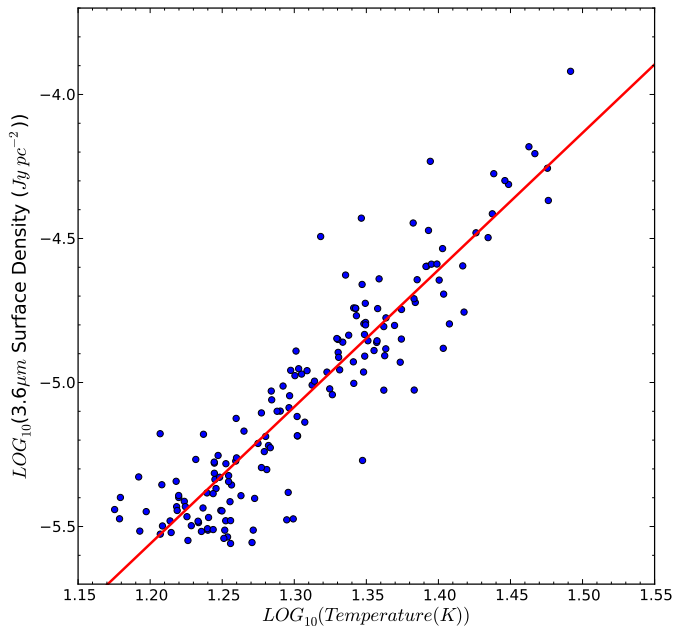}
  \caption{Log(3.6\micron ) flux versus log(Temperature) for the inner 3.1\,kpc. The best-fit linear model is shown
           by the red line.}
  \label{fig:FTcorr}
\end{figure}

The strongest correlation is seen between the dust temperature and the 3.6\micron\ flux in the inner region.
Figure \ref{fig:FTcorr} shows a log-log graph of the two quantities and a linear fit to the points.
The gradient represents the power \textit{n} where $F_{3.6\mu m} \propto T^{n}_{dust}$ where $n=4.61\pm 0.15$. 
While the correlation suggests the dust in the bulge is heated by the general stellar population,
for a modified blackbody with $\beta = 2$ we would expect a gradient of 6. The difference is probably
explained by the simplicity of our assumptions: that there is only a single stellar population in the bulge
and that the bulge has a constant depth in the line of sight. If these assumptions are incorrect, the
3.6\micron\ surface brightness of M31 will only be an imperfect tracer of the intensity of the interstellar radiation field (ISRF).

\begin{figure*}
  \centering
  \includegraphics[trim=0mm 0mm 0mm 0mm, clip=true, width=1.0\textwidth]{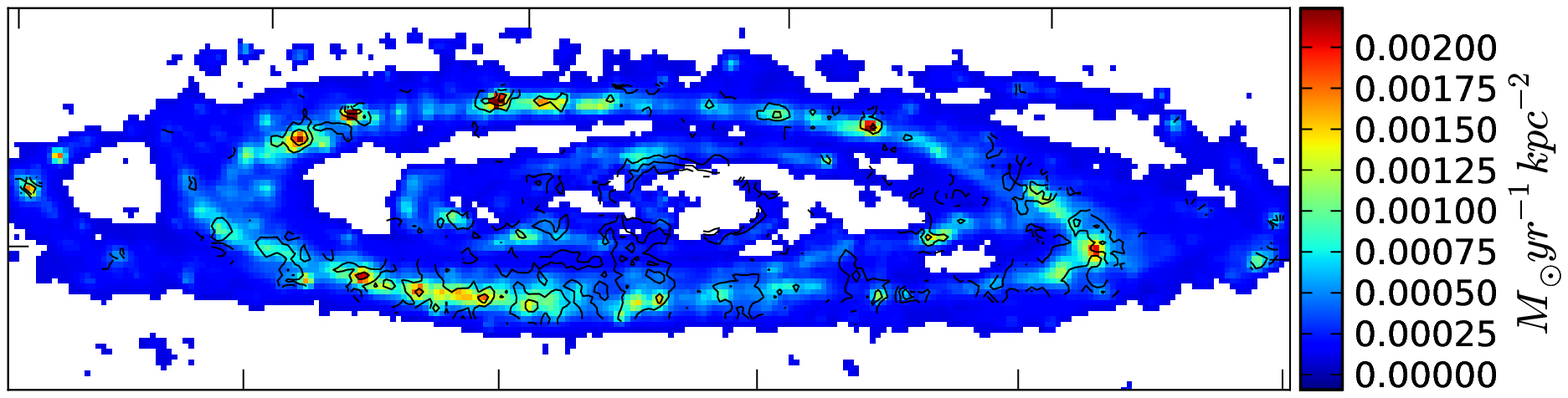}
  \caption{The color image shows the SFR image from G. P. Ford et al. (in preparation) which has been smoothed and re-gridded to match the maps presented in 
           Figure \ref{fig:SED-maps}. The contours are from the dust temperature map and drawn at 18.0, 19.5 and 21.0\,K values.}
  \label{fig:SFRcont}
\end{figure*}

Looking at the temperature beyond 3.1\,kpc, we find a weak, but still highly significant, correlation with SFR, 
suggesting that the ISRF has a significant
contribution from star-forming regions. As most of the star formation in M31 occurs in the
10\,kpc ring this is to be expected. This correlation can be seen in Figure \ref{fig:SFRcont} where 
most but not all of the temperature peaks in the 10\,kpc ring appear aligned with the peaks in the SFR map.
In the same region there is a slight anti-correlation of temperature with the 3.6\micron\ flux which could be explained
by the radial decrease in the 3.6\micron\ flux whilst the dust temperature increases slightly with radius.
The fact that dust temperature increases slightly with radius, while the number-density of stars, traced by
the 3.6\micron, is falling with radius suggests that outside the bulge the dust is mainly heated by young stars.
Nevertheless, the lack of a strong correlation between dust temperature and either SFR or 3.6\micron\ flux suggest the optical/UV light
absorbed by a dust grain is from photons from a large range of distances (e.g., photons from the bulge heating dust in the disk). 

\citet{Bendo2012} have studied FIR colour ratios in M81, M83 and NGC4203. They find the 250/350\micron\ color
ratio has the strongest positive correlation with 1.6\micron\ emission. An increase in the 250/350\micron\ ratio would indicate either 
an increase in dust temperature or a decrease of $\beta$. \citet{Bendo2012} conclude that the most likely explaination is the temperature effect, with 
the dust being heated by the general stellar population traced by the 1.6\micron\ emission. We find a similar correlation to \citet{Bendo2012}, only we trace the
stellar radiation field with the 3.6\micron\ band. However, our SED-fitting results suggest that this is caused by a combination of changes in temperature
and $\beta$. Note that since \citet{Bendo2012} only use color ratios they are unable to discriminate between changes of temperature and $\beta$.
At radii greater than 3.1\,kpc our results suggest the variation in the 250/350\micron\ is mainly caused by a change
in $\beta$. \citet{Bendo2012} 
found that the 70/160\micron\ color ratio has the greatest correlation with star-formation rate, which is evidence that there is dust at more
than one temperature contributing to the 70--500\micron\ emission.
One possible explanation of our failure to find a correlation between the 3.6\micron\ emission and dust temperature outside 3.1\,kpc,
instead of the negative correlation between 3.6\micron\ and $\beta$ we find, might be if the 
\Hersc\ emission at short wavelengths contains a contribution from a warmer dust component. This would mean our fits of a one-component modified blackbody
would produce misleading results.
However, as stated in Section \ref{sect:SEDfit} when we attempted the same
SED fitting process but using all flux densities $\leq$160\micron\ as upper limits, we see little difference in our results.

\subsection{Dust Emissivity and Temperature Relation}
\label{sec:beta-T}

\begin{figure*}
   \centering 
   \includegraphics[trim=12mm 7mm 8mm 6mm,clip=true,width=0.65\textwidth]{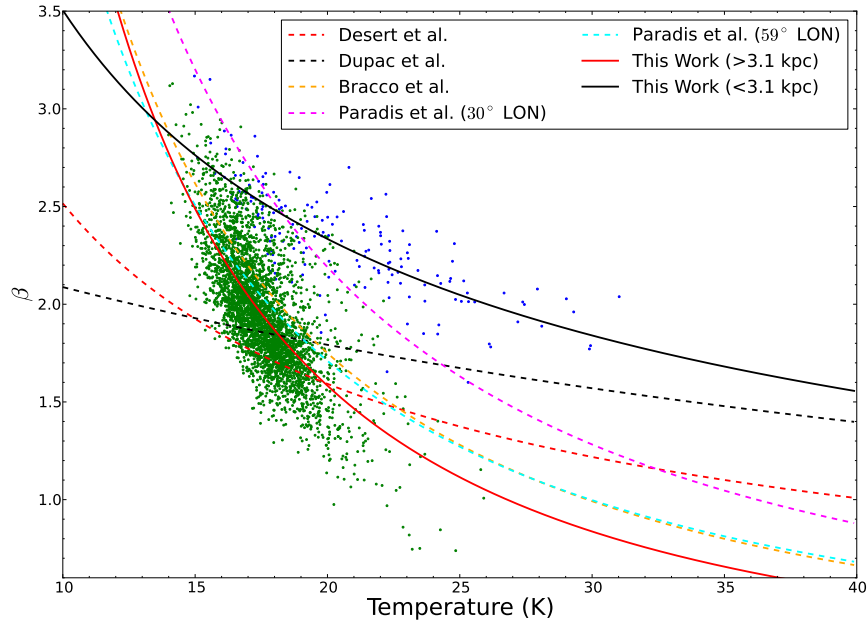}
   \caption{The variation of the dust temperature with emissivity index across M31.  Data points are colour-coded 
            for those within $R<3.1\,\rm kpc$ (blue) and those beyond this radius (green).  Solid lines show the 
            best-fit relations for $T_d-\beta$ in M31. The $T_d-\beta$ relationships in the literature are indicated 
            by the dashed lines \citep[including][]{Dupac2003, Desert2008, Paradis2010, Bracco2011}.}
   \label{fig:B-Tmodels}
\end{figure*}

The dust emissivity index ($\beta$) is related to the physical properties of the dust grains, 
including the grain composition, grain size, the nature of the absorption process and the equilibrium temperature of the dust.  
We would also expect to see a change in $\beta$ due to environment from the processing of the grains via grain 
growth (e.g. coagulation, mantle accretion) or destruction through surface sputtering by ions/atoms or shattering by shocks.
In M31, we detect an apparent inverse correlation between $\rm T_{d}$ and $\beta$ for the inner and outer regions of M31, as shown
in Figure \ref{fig:B-Tmodels}. We find the form of the relation is different for the two regions.

Such an inverse relationship has been observed in the Milky Way with 
previous FIR-submm experiments and surveys including ARCHEOPS \citep{Desert2008},
which showed $\beta$ ranging from 4 to 1 with the dust temperature varying between 7 and 27\,K, and 
PRONAOS \citep{Dupac2003}, which shows a variation of $\beta$ from 2.4--0.8 for dust temperatures between 11 and 80\,K.  
\citet{Veneziani2010} used IR-mm data of Galactic high latitude clouds and found a similar trend, 
and more recently \citet{Paradis2010} with \Hersc\ found a similar inverse relationship with $\beta = 2.7-1.8$ for $\rm T_d=\rm 14-21\,K$ 
for galactic longitude $59^{\circ}$ (at longitude $30^{\circ}$ $\beta = 2.6-1.9$ for 18--23\,K).  
Recently \citet{Bracco2011} used \Hersc-ATLAS observations to investigate $\beta$ variations in 
low-density, high-latitude galactic cirrus, measuring values of $\beta$ ranging from 4.5--1.0 for 
$\rm 10<T_d <28\,K$. 
These $T_d-\beta$ relationships could be indicative of a 
problem with the temperature--$\beta$ degeneracy arising from the SED fitting, the presence of 
dust with a range of temperatures along the line of sight \citep{Shetty2009}, or real variations of the 
properties of the dust grains.

\begin{figure*}
   \centering 
   \includegraphics[trim=12mm 7mm 8mm 6mm,clip=true,width=0.65\textwidth]{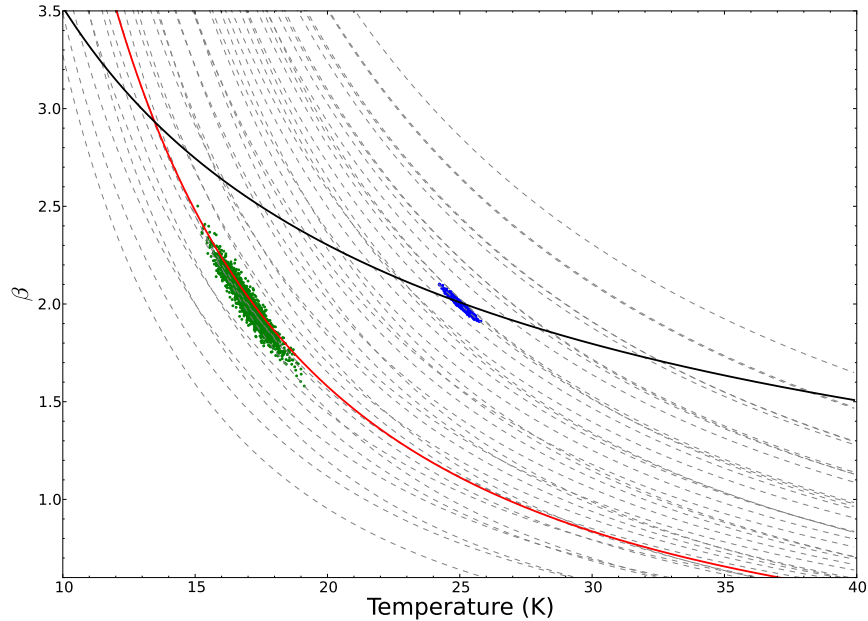}
   \caption{The variation of the dust temperature with emissivity index that arise from just the uncertainties in the 
            measurements. The data shown use the simulations of the SED-fitting method, described
            in Section \ref{sec:simulation}. The green and blue data points show the recovered values of $\beta$ and $\rm T_{d}$
	    for an input model with $\rm T = 17.0$\,K, $\rm \beta = 2.0$ (green) and $\rm T = 25.0$\,K, $\rm \beta = 2.0$ (blue).
            We have carried out the same simulation for input values of $\rm T_{d}$ and $\beta$ over the range
            $\rm T_{d}$ of 15--29\,K in 2\,K intervals and in $\beta$ of 1.6--2.4 in 0.2 intervals. 
            For each group of points we have fitted a line $\rm T_{d} \propto \beta^{n}$, which are the grey dashed lines. 
            In these cases we have not shown the recovered values of $\rm T_{d}$ and $\beta$, merely the lines that are the
            best fit to the points. The red and black solid lines
            are the best fit models to the real data as shown in Figure \ref{fig:B-Tmodels}. When compared with Figure \ref{fig:B-Tmodels}
            it is clear that the uncertainties cannot account for the distribution in the real data.}
   \label{fig:B-Tsim}
\end{figure*}

On the assumption that the inverse correlations between $\beta$ and $\rm T_{d}$ in Figure \ref{fig:B-Tmodels}
are not simply caused by the two variables being separately correlated with radius, we looked for other possible 
causes of the relationships:

\begin{enumerate}
  \item The fitting can lead to a spurious inverse correlation between $\beta$ and $\rm T_{d}$ \citep[][Section \ref{sec:simulation}]{Shetty2009bt}. 
  The most striking feature in Figure \ref{fig:B-Tmodels} is the clear separation in points between the inner 3.1\,kpc and the outer regions.
  To test whether these different distributions might be produced by the fitting artifact, we used the Monte-Carlo simulations from 
  Section \ref{sec:simulation} to simulate the effect of fitting a modified blackbody for various combinations of $\beta$ and $\rm T_d$.
  The grey lines in Figure \ref{fig:B-Tsim} show the best-fit 
  relationships for different input $\rm T_{d}$ and $\beta$ combinations and clearly show that the two different relationships in the two regions cannot be obtained from a
  single population of dust grains. The green and blue data points represent the range of output $\rm T_{d}$ and $\beta$ for an input modified blackbody with 
  $\rm T = 17.0$\,K (green) and $\rm T = 25.0$\,K (blue), with $\rm \beta = 2.0$. A comparison of Figures \ref{fig:B-Tmodels} and \ref{fig:B-Tsim} shows that in both regions 
  of M31 there is a larger range of temperature and dust emissivity for the real data than that found in the Monte-Carlo simulation, 
  indicating there are genuine variations of $\rm T_{d}$ and $\beta$ in both regions. 
  Moreover, the fitting artifact cannot explain the observed relationships of $\beta$  and temperature with radius (Figure \ref{fig:sed-rad}). 

  \item Artificial inverse $\rm T_{d} - \beta$ relationships can also be produced if a one-component modified blackbody model is used to fit dust which contains
  a range of dust temperatures \citep{Shetty2009}. Since we are averaging through the disk of a galaxy along the line of sight (LOS), it is
  obviously possible that the dust contains a range of dust temperatures. While we cannot fully address
  this issue, our $\beta$ values are higher than expected, which is the opposite of what happens from a LOS averaging of temperatures. 
  We also find no statistical evidence from our fits that there is more than one component of dust.
  Also \citet{Paradis2010} and \citet{Anderson2010} show inverse $T_d-\beta$ relationships still exist in places where it is unlikely 
  there is dust at more than one temperature. 

  \item Variation of $\beta$ with wavelength has been reported by some authors both from theoretical models 
  and laboratory experiments and from observations \citep[e.g.,][]{Meny2007, Coupeaud2011},
  with a transition around 500\micron. In Section \ref{sec:excess} we show that there is no evidence for excess 500\micron\ emission, suggesting
  this is not an explanation of our results.
\end{enumerate}

To describe the $T_d-\beta$ relationships we use an empirical model of the form $\beta = AT^{\alpha}$, commonly used
in the literature \citep[e.g.,][]{Desert2008, Paradis2010, PlanckCloud} to fit the $T_d-\beta$ anti-correlation. The observed anti-correlation
between T and $\beta$ may arise due to a change in the physical properties of grains including the grain optical constants changing with
temperature for amorphous grains or changes in the dust emissivity with wavelength \citep[see][and references therein]{Ysard2012}. Other
possibilities include grain growth or quantum mechanical effects (though these latter grain properties only arise at lower dust temperatures than observed in M31).
The best-fit relationship which describes $T_d-\beta$ for R\,$< 3.1\,\rm kpc$ and 
for $\rm 3.1 < R < 15\,kpc$ are (shown in Figure \ref{fig:B-Tmodels}):
\renewcommand{\arraystretch}{1.5}
\begin{equation}
\beta = \left\{
\begin{array}{l l}
  2.30  (\frac{T_d}{20})^{-0.61} & \quad {\rm R < 3.1\,kpc}\\
  1.58 (\frac{T_d}{20})^{-1.57} & \quad {\rm 3.1 \le R < 15\,kpc}\\
\end{array}\right.
\label{eq:varyt}
\end{equation}
\renewcommand{\arraystretch}{1.0}
where the steeper $\rm T_d-\beta$ relationship at R\,$>3.1\,\rm kpc$ agrees well with the relationship found in the plane of the Milky Way at 
longitudes of $59^{\circ}$ \citep{Paradis2010} and in low-density, high-latitude cirrus \citep{Bracco2011}.  
There is some evidence that the $\rm T_d-\beta$ relationship in M31 is slightly steeper, so that for the same temperature compared to the galactic plane, 
M31 has a lower $\beta$ (but this is only at $\sim 5-10\%$ level).

What could be a physical (or chemical) explanation of the different $\rm \beta - T_{d}$ relationships
in the two regions? 
Typical values of $\beta$ are in the range 1.5--2.0 for interstellar dust grains, and have been found in
global extra-galactic studies \citep[e.g.,][]{Skibba2011, Smith2012, Dunne2011} and {\em average} 
global values measured in the Milky Way \cite[e.g.,][]{Paradis2010, Bracco2011, PlanckISM}.
Low values of $\beta$ for large grains would typically represent freshly-formed dust grains in circumstellar disks or 
stellar winds. Alternatively, $\beta \sim 1$ has been observed in regions where small grains dominate \citep{Seki1980}.
High values of $\beta$ ($>2$) might occur due to grain 
coagulation or to the growth of icy mantles on the surface of the grains in denser regions \citep{Aannestad1975, Lis1998, Stepnik2003}.
Studies have also suggested high values of $\beta$ are associated with very cold dust \cite[$\rm T<12\,K$ e.g.,][]{Desert2008}
possibly caused by a change in the absorption properties due to quantum effects at low temperatures,
increasing self-absorption in amorphous grains via tunnelling \citep{Agladze1996, Mennella1998, Meny2007, Paradis2012}.

The highest value of $\beta$ is seen at the 3.1\,kpc boundary between the two regions. This cannot be caused by changes in the
quantum mechanical absorption, since this is only thought to be important for cold dust at temperatures $<12\,\rm K$.
The high $\beta$ values could be due to efficient grain coagulation or mantle growth in dense molecular clouds, although this too
seems unlikely as little CO($J=1-0$) is observed in this region. While there is no obvious explanation for the high $\beta$ values
at this radius, there are many indications that this 3\,kpc ``boundary'' is an interesting regime, we discuss this further in Section \ref{sec:transition}.

In the inner region, we suggest that the decrease in $\beta$ with corresponding increase in $\rm T_d$ might be caused 
by the increased intensity of the ISRF.  
Towards the center of M31, we would expect increased sputtering or sublimation of mantles from the increased ISRF,
shown by the increased temperature of the dust and the increased
X-ray emission observed in the center
\citep{Shirey2001}. The lack of gas in the central regions (Figure \ref{fig:gas-rad}) also suggests that dust is less likely to be shielded 
and thus more efficiently sputtered and leading to smaller grain sizes.  

As we mentioned above, a problem with this analysis is that it is difficult to
determine which are the causal relationships. For example, we have argued that the radial variation in $\rm T_d$ is due to
the radial variation in the ISRF. The radial variation in $\beta$ might then be due to a physical relationship
between $\rm  T_d$ and $\beta$ or it might be the case that there is no causal relationship between these parameters but
the radial variation in $\beta$ is caused by a different effect. For example, an interaction between M32 and M31 might have
caused a wave of star formation which has moved out through the galaxy, which might have led (by a number of processes) to 
the radial variations in $\beta$.
Therefore, we can rule out some hypotheses but we cannot
conclusively determine which is the true explanation using this data set.

\subsection{Why a Transition at 3.1\,kpc?}
\label{sec:transition}

Interpreting the transition in dust properties seen  at 3.1\,kpc (Figure \ref{fig:sed-rad}) is difficult. One possible clue comes from previous gas kinematics studies.
\citet{Chemin2009} found that the H{\sc i} rotation curve inside a 4\,kpc radius is warped
with respect to the rest of disk. \citet{Stark1994} suggest the inner H{\sc i} data is consistent
with a bar extended to 3.2\,kpc, while a newer analysis by \citet{Berman2001} explains the H{\sc i} distribution as the result of a triaxial rotating bulge.
\citet{Block2006}, using \Spitzer\ IRAC observations, identified a new inner dust ring with dimensions
of 1.5\,$\times$\,1\,kpc. By using the stellar and gas distributions and from the presence of the 10\,kpc ring, 
they conclude that an almost head-on collision has occurred between M31 and M32 around 210 million years ago. 
This collision could explain the perturbation of the gas observed in the central 4\,kpc. 
These other observations all show that the inner 3\,kpc of M31 is an intriguing region, although it it not clear what are
the causes of the difference in the dust properties.
The perturbation of the gas may have  
lead to the processing of dust grains, or potentially material from M32 could have been deposited after the interaction. The total dust mass 
for the pixels in our selection within 3.1\,kpc 
is $10^{4.2} M_{\odot}$ which is a plausible amount to be deposited as recently dust masses of $\sim$$10^{5} M_{\odot}$ have been reported in Virgo dwarf ellipticals \citep{Grossi2010}.

Another possibility is the dust properties are affected by the differences between conditions in the bulge and disk. 
\citet{Courteau2011} decomposed the luminosity profile of \Spitzer\ IRAC data into a
bulge, disk and halo. From their Figure 16 we can see that our transition radius of $\sim$3\,kpc is approximately where the bulge emission
begins to become a significant fraction of the optical disk emission. Whether the transition in the dust parameters is due to the 
changing contribution to the ISRF from the general stellar population and 
star formation or if there is another influence in the bulge is unknown.

\subsection{Dark Gas and X factor}
\label{sec:dark}

The detection of `dark gas' in the Milky Way was a surprising early result from \textit{Planck} \citep{Planck2011}, obtained 
by combining \IRAS\ 100\micron\ data and the six-band \textit{Planck} data from 350\micron--3\,mm.
The \textit{Planck} team compared the dust optical-depth with the total column density of hydrogen ($N_{H}^{Tot}$), where the optical depth at each wavelength is given by
\begin{equation}
  \tau_{\nu} = \frac{I_{\nu}}{B(\nu,T_{dust})}
  \label{equ:planck}
\end{equation}
where $I_{\nu}$ is the flux density in that band and $B(\nu,T_{dust})$ is the blackbody function. 
They assumed that at low $N_{H}^{Tot}$ the atomic hydrogen dominates over the molecular component while 
at high column density the molecular hydrogen dominates the emission. 
For these two regimes they found a constant gas-to-dust ratio, 
but at intermediate column densities they found an excess of dust compared to the gas. 
This excess is attributed to gas traced by dust but not by the usual H{\sc i} and CO lines, and is found to be the 
equivalent 28\% of the atomic gas or 118\% of the molecular gas. 
The excess dust emission was typically found around molecular clouds, suggesting that the most likely cause is the presence of molecular gas not traced by the CO line.

We attempted the same analysis as the \citet{Planck2011} for M31 using our SED fitting results from Section \ref{sec:res}. Instead of
using Equation \ref{equ:planck}, we compare the column density of gas estimated from the H{\sc i} and CO
with the column density of dust (for convenience we call this $\Sigma_{dust}$). 
We use this parameter as it is calculated with data from all wavelengths, whereas if we used Equation \ref{equ:planck},
small errors in temperature would cause large uncertainties in $\tau_{\nu}$ for wavelengths close 
to the peak of emission.

\begin{figure*}
  \centering
  \includegraphics[trim=4.0mm 8.0mm 7.0mm 5.0mm, clip=true, width=0.8\textwidth]{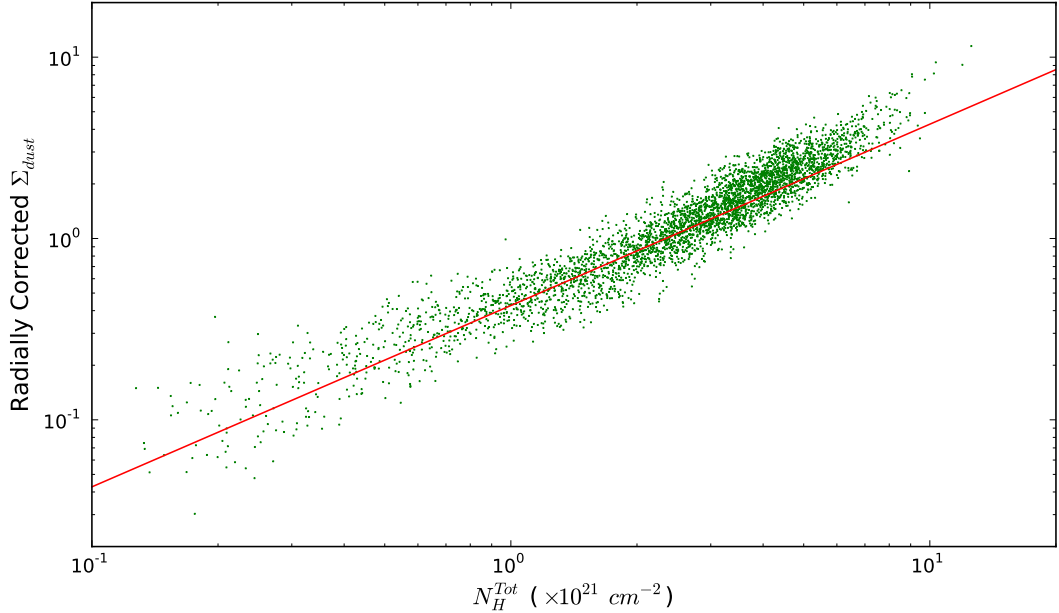}
  \caption{Radially corrected $\Sigma_{dust}$ versus total column density of gas. The plot is shown using our \textbf{best-fit} value of the X-factor of 
           $(1.9 \pm 0.4)\times 10^{20} \rm\ cm^{-2} [K\ kms^{-1}]^{-1}$. The red line represents the best fit model to the data 
           assuming that $\Sigma_{dust} \propto N_H^{Tot}$. The plot shows that, unlike
           the \textit{Planck} data for the Milky Way \citep{Planck2011}, at intermediate gas column densities we do not find an excess
           in dust column density over the best-fit model which would indicate the presence of gas not traced by the H{\sc i} and CO (`dark gas').}
  \label{fig:optdep-gas}
\end{figure*}

\begin{figure*}
  \centering
  \includegraphics[trim=3.0mm 7.0mm 0.0mm 2.0mm, clip=true,width=0.8\textwidth]{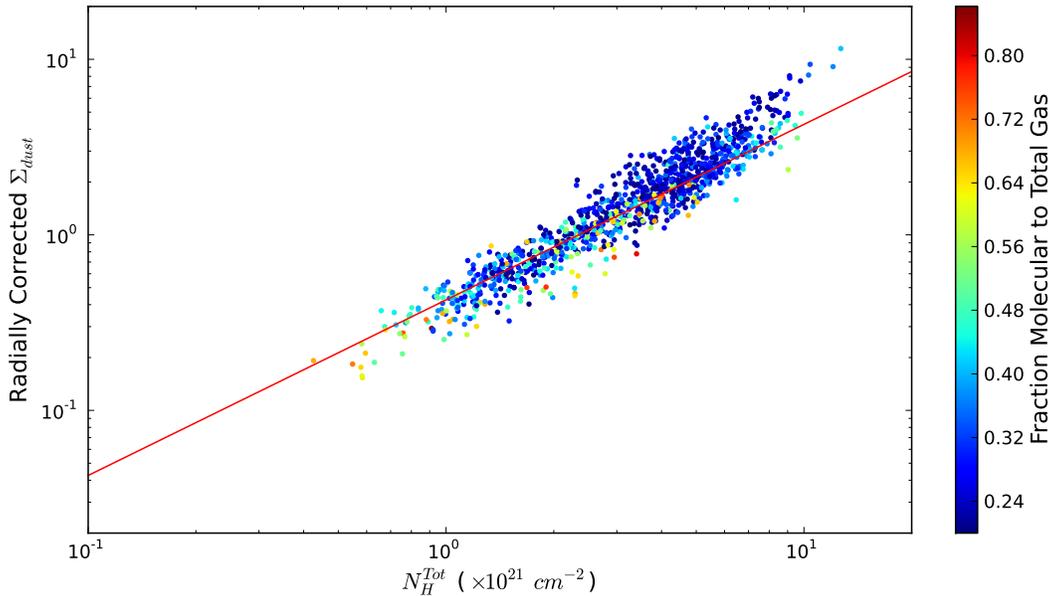}
  \caption{Radially corrected $\Sigma_{dust}$ versus column density of gas for pixels where the molecular fraction
           is greater than 20\%. The data points are colour-coded with the fraction of molecular gas compared to total gas. 
           The Figure shows that the high column densities are not dominated by regions of molecular gas traced by
           the CO. The red line is the fitted model from Figure \ref{fig:optdep-gas}.}
  \label{fig:COdist}
\end{figure*}

The \textit{Planck} team found no radial variation in the gas-to-dust ratio in the Milky Way \citep[see Figure 10][]{Planck212011}.
In Andromeda we show that the gas-to-dust ratio does vary radially (Section \ref{sec:radGD} and Figure \ref{fig:gas-rad}), as expected from the metallicity gradient.
To determine if there is an excess at intermediate column density in dust compared to that expected from the gas we have to correct for the radial change
in gas-to-dust ratio. To remove this dependence, we adjust dust column density by using the exponential fit (shown by the red line in Figure \ref{fig:gas-rad}) 
so the gas-to-dust ratio at all radii has the same value as the center of M31.
To avoid biasing this correction by assuming a value for the X-factor which is highly uncertain,
we estimate this relationship from pixels where the atomic hydrogen column density is $>95$\% of $N_{H}^{Tot}$ (1569 pixels out of the 3600).

\begin{figure*}
  \centering
  \includegraphics[trim=0.0mm 0.0mm 0.0mm 0.0mm, clip=true, width=\textwidth]{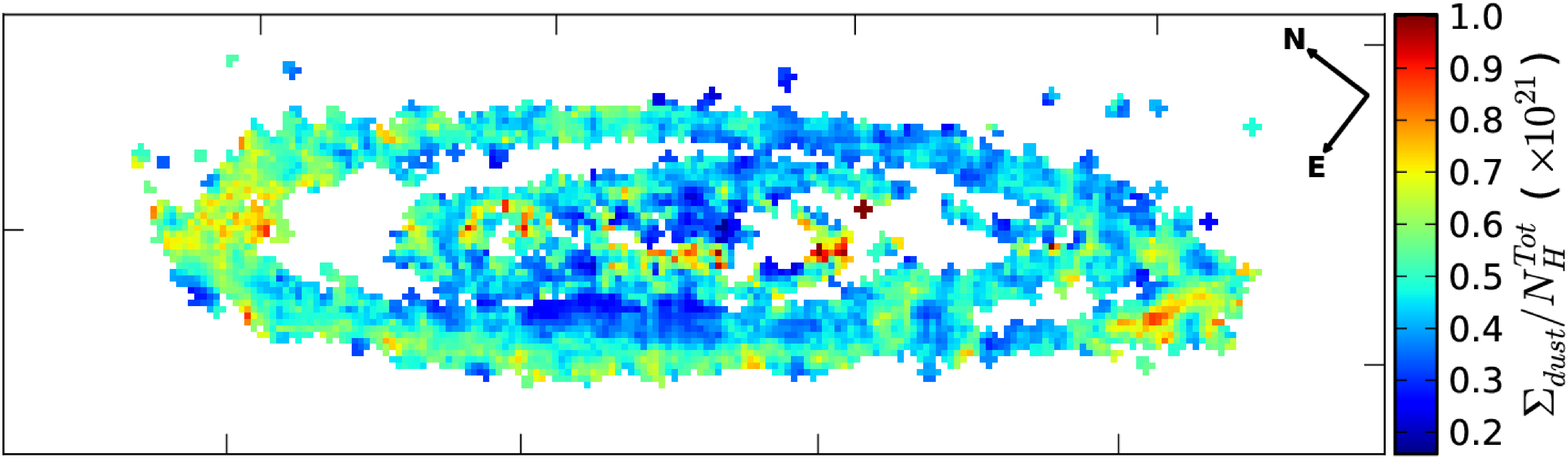}
  \caption{Map of $\Sigma_{dust} / N_{H}^{Tot}$ ratio in M31. Higher values represent areas where there is less gas than 
           predicted from dust measurements. The tick spacing represents 30\arcmin.}
  \label{fig:tauimage}
\end{figure*}

In Figure \ref{fig:optdep-gas} we show that the relationship between corrected dust column density ($\Sigma_{dust}$) and gas column density is well represented by assuming
the two quantities are directly proportional with no excess in dust column-density that could be attributed to `dark gas'. 
Although only a small proportion of the gas is molecular, we still need to use a value for the X-factor. We can estimate this quantity from the data 
itself by finding the values of the X-factor and the constant of proportionality between the gas column density and the corrected $\Sigma_{dust}$ that
gives the minimum $\chi^2$ value (the fitted line is shown in Figure \ref{fig:optdep-gas}.)
We find a best value for the X-factor of $(1.9 \pm 0.4)\times 10^{20} \rm\ cm^{-2} [K\ kms^{-1}]^{-1}$ (or expressed as $\alpha_{CO} = 4.1 \pm 0.9 M_{\odot}\rm \,pc^{-2}\,[K\,kms^{-1}]^{-1}$), 
where the random error is estimated using a monte-carlo technique
(similar to one used in Section \ref{sect:SEDfit}). For the dust column
densities in each pixel we use the uncertainties provided by the SED fitter as explained in Section \ref{sect:SEDfit}, which is on average $\sim$22\%. \citet{Nieten2006}
quote a calibration error of 15\% for the CO observations (which directly results in at least a 15\% uncertainty in the X-factor) 
which we combine with the noise in each pixel of our processed moment-zero CO map. 
For the H{\sc i} observations we use an uncertainty map provided by Robert Braun (private communication) which has an average uncertainty of 12\% on the raw 10\arcsec\ map.
We also include a 5\% systematic uncertainty (e.g., calibration uncertainties).
If the opacity corrected H{\sc i} map is used
our best fit X-factor is $(2.0 \pm 0.4)\times 10^{20} \rm\ cm^{-2} [K\ kms^{-1}]^{-1}$.

\citet{Leroy2011} find values of the X-factor between $0.97$ to $4.6\times 10^{20} \rm\ cm^{-2} [K\ kms^{-1}]^{-1}$ when analysing the southern, northern and inner regions for M31 with \Spitzer\ data. 
Our average value of the X-factor falls within their range of values. A full analysis of the spatial variations of the X-factor with our data will be undertaken in a future paper.

There are two main problems with this method which could lead us to miss `dark gas'. 
First, unlike the Milky Way we have to average through the whole disk of M31 and, second, M31 has a significantly lower molecular gas fraction. 
The latter prevents us from fitting the model to pixels with just very high and low values of $N_{H}^{Tot}$ as the pixels with highest molecular gas fraction are not clustered
to high $N_{H}$ values (this is illustrated in Figure \ref{fig:COdist}). This suggests Andromeda may not be the best galaxy for this
analysis as the molecular contribution to the overall column density is quite low. 

However, we can also try an alternative method of looking for `dark gas' because we have one important advantage over the 
\textit{Planck} team: we can see M31 from the outside. 
We can therefore make a map of the ratio of radially corrected dust column density ($\Sigma_{dust}$) to gas column density to look for regions of enhancements in this ratio 
(Figure \ref{fig:tauimage}). The image clearly shows spatial variations which could either suggest regions of `dark gas' or local variations in the metallicity or emissivity of dust.
To distinguish between these scenarios an independent measurement of the `dark gas' is required. On the outskirts of molecular clouds, CO could be photodissociated and the 
carbon gas would therefore reside in C or C$^+$ \citep{Wolfire2010}.
Planned observations of the C[II] 158\micron\ line in Andromeda with \Hersc\ could then be a potential test for investigating whether `dark gas' exists in M31.

\section{Conclusions}
\label{sec:conc}

In this paper we present the results of an analysis of dust and gas in Andromeda using new \Hersc\ observations from the HELGA survey.
We have $\sim$4000 independent pixels with observations in the range of 70--500\micron. We find the following results:

\begin{enumerate}

\item We find that a variable dust emissivity index, $\beta$, is required to adequately fit all the pixels in Andromeda. When a variable $\beta$ is used,
the modified blackbody model with a single temperature is found to be a statistically reasonable fit to the data in the range 100--500\micron. 
There is no significant evidence of an excess of dust emission at 500\micron\ above our model. 

\item There are two distinct regions with different dust properties, with a transition at $R=3.1$\,kpc. In the center
of Andromeda, the temperature peaks with a value of $\sim$30\,K and a $\beta$ of $\sim$1.9. The temperature then declines radially to a value of 
$\sim$17\,K at 3.1\,kpc with a corresponding increase in $\beta$ to $\sim$2.5. At radii larger than 3.1\,kpc $\beta$ declines but only with a small
associated increase in temperature.

\item The drop in $\beta$ towards the center of the galaxy may be caused by increased sputtering or sublimation of mantles from an increased
ISRF. The origin of the high $\beta$ values at 3.1\,kpc from the center is less clear but may be indicative of either grain coagulation or an increase in the growth of
icy mantles.

\item The dust surface density for our pixels in which flux is detected at $>5\sigma$ in all Herschel bands 
range is between $\sim$0.1--2.0\,$M_{\odot}\rm\ pc^{-2}$. We find the gas-to-dust ratio
increases exponentially with radius. The gradient matches that predicted from the
metallicity gradient assuming a constant fraction of metals are included into dust grains.
The dust surface density is correlated with the star formation rate rather than with the stellar surface density.

\item In the inner 3.1\,kpc the dust temperature is correlated with the 3.6\micron\ flux. This suggests the heating of the dust in the 
bulge is dominated by the general stellar population. Beyond 3.1\,kpc there is a weak correlation
between dust temperature and the star formation rate.

\item We find no evidence for `dark gas', using a similar technique as the \textit{Planck} team. However, we find this technique may not
be as effective for M31 due to poor angular resolution and line of sight effects. We have used an alternative technique by constructing a
gas-to-dust map after correcting for the radial gradient.
We do find regions with enhancements (i.e., higher values of $\Sigma_{dust} / N_H^{Tot}$), which may show places where `dark gas' exists, or may be due
to local variations in the gas-to-dust ratio. 
A detection of a potential component of CO-free molecular gas will be possible with future observations to measure the C[II] line planned with \Hersc.

\item By minimising the scatter between our radially corrected dust column-density and the column-density of gas inferred from the H{\sc i} and CO line 
we find a value for the X-factor of $(1.9 \pm 0.4)\times 10^{20} \rm\ cm^{-2} [K\ kms^{-1}]^{-1}$ (or expressed as $\alpha_{CO} = 4.1 \pm 0.9 M_{\odot}\rm \,pc^{-2}\,[K\,kms^{-1}]^{-1}$).

\end{enumerate}

Our results of Andromeda represent the largest resolved analysis of dust and gas in a single galaxy with \Hersc. 
The results of this analysis on M31 is strikingly different from those obtained by the Planck team in the Milky Way, 
since we find no clear evidence for `dark gas', a radial gradient in the gas-to-dust ratio and evidence for radial
variation in the dust emissivity index ($\beta$). In future work it will be important to understand these differences 
between the two big spirals in the local group.

\acknowledgments We thank everyone involved with the {\it Herschel}
Observatory.  PACS has been developed by a consortium of institutes
led by MPE (Germany) and including UVIE (Austria); KU Leuven, CSL,
IMEC (Belgium); CEA, LAM (France); MPIA (Germany); INAF-
IFSI/OAA/OAP/OAT, LENS, SISSA (Italy); IAC (Spain). This development
has been supported by the funding agencies BMVIT (Austria), ESA-PRODEX
(Belgium), CEA/CNES (France), DLR (Germany), ASI/INAF (Italy), and
CICYT/MCYT (Spain). CVN, PR, DH, GR, YN and KE acknowledge support
from the Belgian Federal Science Policy Office via the PRODEX
Programme of ESA.  

SPIRE has been developed by a consortium of
institutes led by Cardiff Univ. (UK) and including:
Univ. Lethbridge (Canada); NAOC (China); CEA, LAM (France); IFSI,
Univ. Padua (Italy); IAC (Spain); Stockholm Observatory (Sweden);
Imperial College London, RAL, UCL-MSSL, UKATC, Univ. Sussex (UK); and
Caltech, JPL, NHSC, Univ. Colorado (USA). This development has been
supported by national funding agencies: CSA (Canada); NAOC (China);
CEA, CNES, CNRS (France); ASI (Italy); MCINN (Spain); 
SNSB (Sweden); STFC, UKSA (UK); and NASA (USA). 

HIPE is a joint
development by the \Hersc~Science Ground Segment Consortium,
consisting of ESA, the NASA \Hersc~Science Center and the HIFI, PACS
and SPIRE consortia. 

GG is a postdoctoral researcher of the FWO-Vlaanderen (Belgium)

The research leading to these results has received funding from the 
European Community's Seventh Framework Programme (/FP7/2007-2013/) under 
grant agreement No 229517.

{\it Facilities:} \facility{Herschel (PACS and SPIRE)} , \facility{Spitzer (MIPS)}.

\bibliographystyle{apj}

\clearpage
\end{document}